\documentclass[longauth]{aa}

\usepackage{graphicx}
\usepackage[dvipsnames]{xcolor}
\usepackage[utf8]{inputenc}

\usepackage{txfonts}
\usepackage{hyperref}
\usepackage{subfig}
\usepackage{orcidlink}
\usepackage{amssymb}
\usepackage{footnote}
\makesavenoteenv{table}
\makesavenoteenv{tabular}

 \hypersetup{colorlinks=true, citecolor=[RGB]{129, 123, 184},  urlcolor=[RGB]{129, 123, 184},  linkcolor=[RGB]{73, 46, 128},  breaklinks=true}

\defcitealias{Fitzpatrick1999}{F99}     
\defcitealias{boone2021}{B21a}
\defcitealias{boone2021a}{B21b}

\begin{document} 

   \title{ZTF-SEDm Type Ia supernova sample\\for Twins Embedding spectrophotometric standardization}

   \author{Ganot,~C.\inst{\ref{ip2i}}\thanks{\email{c.ganot@ip2i.in2p3.fr}}\orcidlink{0009-0000-6300-9174}
    \and  Copin,~Y.\inst{\ref{ip2i}}\orcidlink{0000-0002-5317-7518} 
    \and Rigault,~M.\inst{\ref{ip2i}}\orcidlink{0000-0002-8121-2560}
    \and Dimitriadis,~G.\inst{\ref{lancaster}}\orcidlink{0000-0001-9494-179X}
     \and Goobar,~A.\inst{\ref{okc}}\orcidlink{0000-0002-4163-4996}
    \and Maguire,~K.\inst{\ref{dublin}}\orcidlink{0000-0002-9770-3508}
    \and Nordin,~J.\inst{\ref{berlin}}\orcidlink{0000-0001-8342-6274}
    \and Smith,~M.\inst{\ref{lancaster}}\orcidlink{0000-0002-3321-1432}
\and Aldering,~G.\inst{\ref{lbnl}}
\and Barjou-Delayre,~C.\inst{\ref{clermont}}\orcidlink{0009-0000-8510-8982}
\and Betoule,~M.\inst{\ref{lpnhe}}\orcidlink{0000-0003-0804-836X}
\and Bloom,~J.~S.\inst{\ref{lbnl},\ref{berkeley_astro}}\orcidlink{0000-0002-7777-216X}
\and Burgaz,~U.\inst{\ref{dublin}}\orcidlink{0000-0003-0126-3999}
\and Galbany,~L.\inst{\ref{ice-csic},\ref{ieec}}\orcidlink{0000-0002-1296-6887}
\and Ginolin,~M.\inst{\ref{cambridge},\ref{ip2i}}\orcidlink{0009-0004-5311-9301}
\and Graham,~M.\inst{\ref{pma}}\orcidlink{0000-0002-3168-0139}
\and Hale,~D.\inst{\ref{caltech}}\orcidlink{0000-0002-4662-122X}
\and Johansson,~J.\inst{\ref{okc}}\orcidlink{0000-0001-5975-290X}
\and Kasliwal,~M.M.\inst{\ref{pma}}\orcidlink{0000-0002-5619-4938}
\and Kim,~Y.-L.\inst{\ref{seoul}} \orcidlink{0000-0002-1031-0796}
\and Masci,~F.J.\inst{\ref{ipac}}\orcidlink{0000-0002-8532-9395}
\and Müller-Bravo,~T.E.\inst{\ref{dublin},\ref{icen}}\orcidlink{0000-0003-3939-7167}
\and Perlmutter,~S.\inst{\ref{lbnl},\ref{berkeley}}
\and Popovic,~B.\inst{\ref{southampton}}\orcidlink{0000-0002-8012-6978}
\and Purdum,~J.N.\inst{\ref{caltech}}\orcidlink{0000-0003-1227-3738}
\and Rusholme,~B.\inst{\ref{ipac}}\orcidlink{0000-0001-7648-4142}
\and Sollerman,~J.\inst{\ref{okc_astro}}\orcidlink{0000-0003-1546-6615}
\and Terwel,~J.H.\inst{\ref{dublin}}\orcidlink{0000-0001-9834-3439}
\and Townsend,~A.\inst{\ref{berlin}}\orcidlink{0000-0001-6343-3362}
 }

   \institute{
   IP2I Lyon/IN2P3, CNRS, Universite Claude Bernard Lyon 1,   UMR 5822, F-69622, Villeurbanne, France
   \label{ip2i}
   \and
   Department of Physics, Lancaster University, Lancaster, LA1 4YB, UK \label{lancaster}
   \and
   The Oskar Klein Centre, Department of Physics, AlbaNova, SE-106 91 Stockholm , Sweden \label{okc} 
   \and
   School of Physics, Trinity College Dublin, College Green, Dublin 2, Ireland \label{dublin}
   \and
        Institut für Physik, Humboldt-Universität zu Berlin, Newtonstr. 15, 12489 Berlin, Germany \label{berlin}
        \and
        Physics Division, Lawrence Berkeley National Laboratory, 1 Cyclotron Road, Berkeley, CA, 94720, US \label{lbnl}
        \and
        Université Clermont Auvergne, CNRS/IN2P3, LPCA, F-63000 Clermont-Ferrand, France \label{clermont}
        \and 
        Sorbonne Université, CNRS/IN2P3, LPNHE, F-75005, Paris, France \label{lpnhe}
        \and 
        Department of Astrophysics, University of California, Berkeley, CA 94720-3411, USA \label{berkeley_astro}
        \and
        Institute of Space Sciences (ICE-CSIC), Campus UAB, Carrer de Can Magrans, s/n, E-08193 Barcelona, Spain. \label{ice-csic}
        \and
        Institut d'Estudis Espacials de Catalunya (IEEC), 08860 Castelldefels (Barcelona), Spain \label{ieec}
        \and
        Institute of Astronomy and Kavli Institute for Cosmology, University of Cambridge, Madingley Road, Cambridge CB3 0HA, UK \label{cambridge}
        \and 
        Division of Physics, Mathematics, and Astronomy, California Institute of Technology, Pasadena, CA 91125, USA
        \label{pma}
        \and
        Caltech Optical Observatories, California Institute of Technology, Pasadena, CA 91125, USA \label{caltech}
        \and
        Department of Astronomy \& Center for Galaxy Evolution Research, Yonsei University, Seoul 03722, Republic of Korea \label{seoul}
        \and
        IPAC, California Institute of Technology, 1200 E. California Blvd, Pasadena, CA 91125, USA \label{ipac}
        \and
        Instituto de Ciencias Exactas y Naturales (ICEN), Universidad Arturo Prat, Chile \label{icen}
        \and
        Department of Physics, University of California, Berkeley, 366 Physics North, MC 7300, Berkeley, CA 94720, USA \label{berkeley}
        \and 
        University of Southampton, Highfield, Southampton, England, UK \label{southampton}
        \and
        The Oskar Klein Centre, Department of Astronomy, AlbaNova, SE-106 91 Stockholm , Sweden \label{okc_astro} 
        }
 
  \abstract
   {}
   {This paper has two aims: the first aim is to build a large homogeneous spectrophotometric sample of Type Ia supernovae (SNe~Ia) from the second data release of the \textit{Zwicky} Transient Facility (ZTF DR2). We used the spectrum sample from the low-resolution ($R\sim100$) SEDmachine (SEDm) Integral Field Spectrograph (IFS) that gathered 3069 spectra. This is one of the largest samples of such collections that can attempt to reproduce the Twins Embedding (TE) spectrophotometric standardization method. This is our second objective.
   The method was developed based on high-quality spectra from 200 SNe~Ia of the Nearby Supernova factory (SNfactory) and led to an exceptionally low value of 0.073 mag for the intrinsic scatter. }
   {As the SEDm is not designed as a spectrophotometric instrument, we first improved the flux-calibration accuracy of the SN~Ia spectrum sample using the ZTF photometric data, which were calibrated at the percent level. We corrected the spectra for second-order polynomials, fitted by comparing the synthetic photometry in the ZTF $g$, $r$, $i$ filters with the light-curve (LC) data. We then applied the three steps of the TE parameterization to a subset of 783 ZTF SN spectra near maximum light while comparing results from SNfactory and ZTF. We finally analyzed the standardization methods based on the TE parameters.
    }
   {The precision of the phase-correction model, which is the first step of the TE, is estimated at 0.01~mag in $g$ band based on ZTF data. Despite the challenge posed by the spectrum-extraction pipeline associated with the SEDm (flux calibration, leftover host signal, low signal-to-noise ratio, and low resolution), we applied a first standardization in color based on the second step of the TE, called read between the lines (RBTL), to the ZTF sample. We reached a Hubble residual scatter of 0.153~mag, all in normalized median absolute deviation, which is to be compared to the $\sim0.11$~mag obtained with the SNfactory data. The SALT color and stretch standardization reaches a scatter of 0.164~mag for the same ZTF SN~Ia sample, and its host steps are $\sim$0.1~mag and zero for RBTL. When considering the scatter due to the redshift error and flux calibration error, we estimated a RBTL scatter of $\sim0.129$~mag for this ZTF sample as an upper limit because we identified an additional contribution from a systematic error in color. We tested the standardization based on the nonlinear TE parameters, and, as expected from the low spectrum quality, it did not improve the overall dispersion.}
   {We release 1897 flux calibrated spectra of 1607 SNe~Ia with an estimated photometric accuracy of 0.07~mag. We further demonstrate that some amount of spectrophotometric SN~Ia standardization can be done with limited-quality spectra. The RBTL standardization is more efficient than that of SALT with one parameter less, and the resulting host steps are consistent with zero. This makes it less prone to astrophysical bias. 
   For future spectroscopic surveys, targeting the extraction pipeline for a thorough flux calibration and good signal-to-noise ratio would enable us to compute the full TE standardization, which would further reduce the scatter in the distance estimate.
   }

   \maketitle

\section{Introduction}
\label{sec:introduction}

 Supernovae~Ia (SNe Ia) are standardizable cosmological candles that are used to derive the expansion history of the Universe. They allowed the discovery that the expansion of the Universe accelerates, which is thought to be caused by an unknown dark energy \citep{riess1998a, perlmutter1999a}. Since then, cosmologists have focused on measuring the properties of dark energy, probing its equation of state parameter $w_0$ and its potential evolution with cosmic time $w_a$. If dark energy is a simple cosmological constant $\Lambda$ in Einstein's equation of gravitation, we would expect $w_0=-1$ at all times (hence, $w_a=0$). However, recent combinations of baryon acoustic oscillation data, anisotropies in the cosmic microwave background, and the SNe~Ia distance-redshift relation have led to a $\sim4\sigma$ tension with $\Lambda$, finding $w_0=-0.727\pm0.067$ and $w_a=-1.05^{+0.31}_{-0.27}$ \citep{adame2025, planck2020,des2024}. 

Data from SN~Ia are the most sensitive probe of the recent expansion of the Universe dominated by dark energy \citep{efstathiou2025}. The tension varies from $\sim2.5\sigma$ to $\sim4\sigma$ depending on the choice of the SN~Ia dataset and the cosmological inference pipeline \citep{brout2022, rubin2025, des2024, popovic2025a}.

This new tension in $\Lambda$ adds to the nearly decade-old Hubble tension. The Hubble constant, $H_0$, is found to have a higher value, at $5\sigma$, when derived by calibrating the absolute magnitudes of SNe~Ia from a direct distance ladder, in comparison to the expectation of the standard model of cosmology, whose parameters are anchored by early-epoch data \citep{riess2016,riess2024,planck2020}.

If they were confirmed, these tensions would prove the current standard model of cosmology to be incorrect, and new fundamental physics should emerge. The physical nature of SNe~Ia, however, which are the key cosmological probes for both of these results, is not fully understood so far, and deriving distances from SN~Ia data is ultimately an empirical procedure. This affects the accuracy of the derived distances, and therefore, of the inferred cosmology.

The peak magnitudes of SN~Ia have a natural scatter of $\sim0.4$~mag, which is reduced to $\sim0.15$~mag when two empirical relations, based on imaging data, are taken into account. These two relations connect their peak magnitudes with properties of their light curves (LCs): SNe~Ia with fast-evolving LCs are fainter, as are those with redder colors. These are usually referred to as the brighter-slower and redder-fainter relations \citep{phillips1993, tripp1998,riess1996a}. The two parameters that are involved can be estimated by fitting spectrophotometric templates on broadband photometric LCs, that is, using the reference SALT model \citep{guy2007, betoule2014}. 

When these two relations are accounted for, the residual dispersion is dominated by an intrinsic unexplained scatter of $\sim~0.1$~mag (e.g.,~\citealt{betoule2014,scolnic2018a}), and the standardized SN~Ia magnitudes are correlated with the properties of their environment. This can either be global, such as the host stellar mass \citep[e.g.,][]{sullivan2010, kelly2010, childress2013, rigault2020, smith2020}, or local to the SN~Ia, such as the local color or star formation rate \citep[e.g.,][]{rigault2013, roman2018a, jones2018, rigault2020, kelsey2021}. It raises questions about the accuracy of this photometric SN~Ia standardization \citep{ginolin2025, ginolin2025a,popovic2025}, and consequently, of the derived distances and cosmological parameters.

Alternatively, \cite{bailey2009a} first demonstrated that we can standardize SNe~Ia using spectrophotometric information alone, with a precision that is competitive with that obtained using photometry. \cite{fakhouri2015} then introduced the concept of twins, demonstrating based on spectrophotometric timeseries data from the Nearby Supernova factory~\citep[SNfactory or SNf,][]{aldering2002}, that two different SNe~Ia can be extremely alike spectroscopically. The authors suggested that these twins went through the same explosion mechanism, which is thus expected to correspond to a similar absolute brightness. The authors indeed showed that by grouping SNe~Ia by spectral similarity, a scatter along the Hubble diagram as low as $\sim0.07$~mag can be reached. This leaves much less room for unaccounted scatter or astrophysical biases. The authors demonstrated that using a single spectrum at maximum light is enough to reach this precision. This method would allow us to dispense with photometry when the SN is found and the phase is known, and to only use spectra to infer SN~Ia distances.

This initial study was expanded into a formal parameterization of twins using an enlarged dataset, and it was developed as a standardization method by \cite{boone2021, boone2021a} (hereafter \citetalias{boone2021} and \citetalias{boone2021a}). The Twins Embedding (TE) is a three-parameter nonlinear space that captures the SN~Ia spectral diversity at maximum light after accounting for the estimated reddening from spectral domains that are far from strong and highly variable features. These three parameters along with the estimated color can then be used to standardize SN~Ia peak magnitudes, leading to a similar scatter of 0.073~mag. More recently, \cite{stein2022} used a neural network probabilistic auto-encoder to extend this work using the same dataset. They employed SNe~Ia with sparse spectrophotometric time series without the requirement of a spectrum near peak luminosity.

While highly encouraging, these results have been limited to the spectrophotometric time series of the SNfactory dataset so far. We here investigate whether these results can be recovered using another spectral dataset, and we focus on the TE method to do this. We used the recently published ZTF SN~Ia DR2 sample \citep{rigault2025}, which gathered 3628 spectroscopically confirmed SNe~Ia acquired between March 2018 and December 2020 by the ZTF survey \citep{bellm2019, graham2019}.

This paper is structured as follows: we first present the characteristics of the ZTF SN~Ia DR2, in particular, the SEDm spectroscopic sample, in Sect.~\ref{sec:ztf_data} . In this section, we improve the flux calibration of the SEDm spectra based on the associated photometric data. We compare the resulting spectrophotometric sample to that of SNfactory, based on which the TE standardization method was developed.

Second, we summarize the TE standardization method in Sect.~\ref{sec:twins_embedding_method}, and we present the ZTF SN~Ia subset defined for the TE application in Sect.~\ref{sec:ztf_vs_snf} . We apply the first step of the TE, which converts it into a spectrum sample at maximum light. Third, we present the results of the TE spectrophotometric standardization on ZTF: the method called read between the lines (hereafter RBTL) in Sect.~\ref{sec:RBTL} and the full TE in Sect.~\ref{sec:te}. We discuss these results in Sect.~\ref{sec:discussion} and evaluate the dependence of the standardization method on several SN parameters and their environment. We conclude in Sect.~\ref{sec:conclusion}.

We use the AB-magnitude system and keep the parameter notation from \citetalias{boone2021}, except for the TE color parameter $\Delta \tilde{A}_V$, which we simplify to $\Delta {A}_{V}$ for clarity. The dispersions are in nMAD when not specified otherwise.

\section{Building the ZTF spectrophotometric dataset}\label{sec:ztf_data}

As the ZTF SN Ia DR2 is the largest homogeneous SN Ia dataset to date \citep{rigault2025}, we wished to take advantage of this large ZTF spectrum sample to apply the spectrophotometric standardization method to the possible extent. The primary objective of the ZTF spectrograph is follow-up and spectral typing, and its dataset quality was not initially sufficient for this study because of the non-spectrophotometric extraction pipeline and residual contaminations.

We exploited the calibrated photometric data to build a spectrophotometric sample, which is essential for deriving distances. We examined the limitations of the flux calibration method due to host signal residuals and low SNR of the sample by quantifying the calibration precision. Section~\ref{sec:ztfsniadr2} describes the ZTF DR2 SN~Ia dataset, Sect.~\ref{sec:calib} introduces the flux calibration method and estimates its accuracy and limitations, and Sect.~\ref{subsec:pres_sample} presents the spectrophotometric sample of 1897 spectra from 1607 SNe Ia that we release.

\subsection{ZTF spectroscopic sample}
\label{sec:ztfsniadr2}

The ZTF \citep{bellm2019, graham2019, masci2019, dekany2020a} is a time-domain survey that started scientific operation in March 2018 and is expected to continue until the end of 2026. The ZTF camera is mounted on the 48-inch Schmidt telescope at Mount Palomar Observatory and monitors the northern sky with a typical cadence of one to three days in the $g$, $r$, and $i$ optical bands. The 47 $\deg^2$ wide camera reaches a typical depth of 20.5~mag in $g$ and $r$ and of 20~mag in $i$ in 30 seconds. 

The ZTF survey has dedicated access to the SEDm \citep[SEDm;][]{blagorodnova2018, rigault2019}, a low-resolution ($R\sim100$) integral field spectrograph covering the 3776--9223~\AA{} wavelength range that was designed to spectrally classify SN-like transients detected by ZTF. In 30 minutes, the SEDm can acquire spectra with a precision sufficient for a classification of SN type up to $\sim$19~mag in $r$ band. 
The ZTF collaboration also has access to other spectroscopic facilities through specific follow-up proposals.

As part of the Bright Transient Survey \citep[BTS;][]{fremling2020} program, the ZTF collaboration used these spectroscopic facilities to classify all SN-like transients up to 18.5~mag and the majority of those up to 19~mag \citep{perley2020}. Of these SNe, $\sim70\%$ are of Type Ia. The ZTF SN~Ia DR2 dataset of 3628 SNe~Ia largely overlaps with the BTS sample and has 80\% of the sources in common \citep{rigault2025}. The other sources are fainter \citep{smith2025}. As a consequence, the ZTF SN Ia DR2 sample is free of nonrandom selection (i.e., volume-limited) up to $z\leq0.06$ \citep{amenouche2025, rigault2025}.

Every SN Ia in the ZTF SN Ia DR2 release has at least one spectrum that led to a secured classification for a total of 5138 spectra, 3069 of which come from the SEDm. Of these SNe~Ia, 2630 are non-peculiar SNe~Ia with good sampling and pass the usual cosmological quality cuts.  The ZTF SN~Ia DR2 contains 3652 spectra associated with these SNe~Ia, 2273 (64\%) of which come from the SEDm. 

In this first analysis, in which we study the application of TE on a non-SNfactory dataset, we focused on this homogeneous sample of 2273 SEDm spectra from 1872 SNe~Ia. 
These spectra were extracted using the data-processing procedure described in \cite{rigault2019}, with a typical precision of the wavelength solution of 3~\AA{} and an accuracy in the color calibration of $\sim5\%$. The relative and absolute spectrophotometry of the sample is unknown a priori.

The background host contamination was handled using the spectroscopic scene modeling \texttt{Hypergal}, described in \cite{lezmy2022a}. In this pipeline, a pure-host cube was estimated using SED fitting on public optical photometric images. This host-cube model was then aligned to match actual observing conditions on a given SEDm exposure. The SN spectrum was obtained by using this host-cube model as a background; the whole fit was made using a single forward-modeling procedure. Leftover narrow wavelength features that are not well constrained by the SED fit might still contaminate the extracted SN spectrum, however, as these features are hard to model properly based on only a few photometric images. Furthermore, targets within the core of the host still show a strong background signal despite the \texttt{Hypergal} procedure (see details in \citealt{lezmy2022a}).

\subsection{Flux calibration of SEDm spectra}
\label{sec:calib}

The absolute flux calibration of SEDm spectra is unknown because it was not necessary for the SN typing process. To recover it and correct for the spectral zero point (ZP($\lambda$)) of the flux calibration, we compared the spectra to LC data to set a low-order ZP($\lambda$). We present the flux calibration method and estimate the precision of the flux calibration.

\subsubsection{Flux calibration method}
\label{subsec:calib_method}

Each SN of the spectroscopic sample was initially observed with the ZTF camera. SALT2.4 \citep{guy2007, betoule2014} interpolates the LC data points in time, and the resulting LCs are flux calibrated to the percent level \citep{rigault2025}. As the photometric data have a good coverage in time in $r$ and $g$, the interpolation error in time is marginal. About $19\%$ of the SNe have no data in $i$, and the uncertainty resulting from the extrapolation to this filter is estimated by SALT. The light-curve amplitudes $x_0$, computed from ZTF LCs, have an absolute floating ZP known at 9\%, but a relative calibration of 1\% \citep{lacroix2025}. We therefore expect a variation of 1\% from the LCs calibration.

\begin{figure}
    \centering
    \includegraphics[width=19pc]{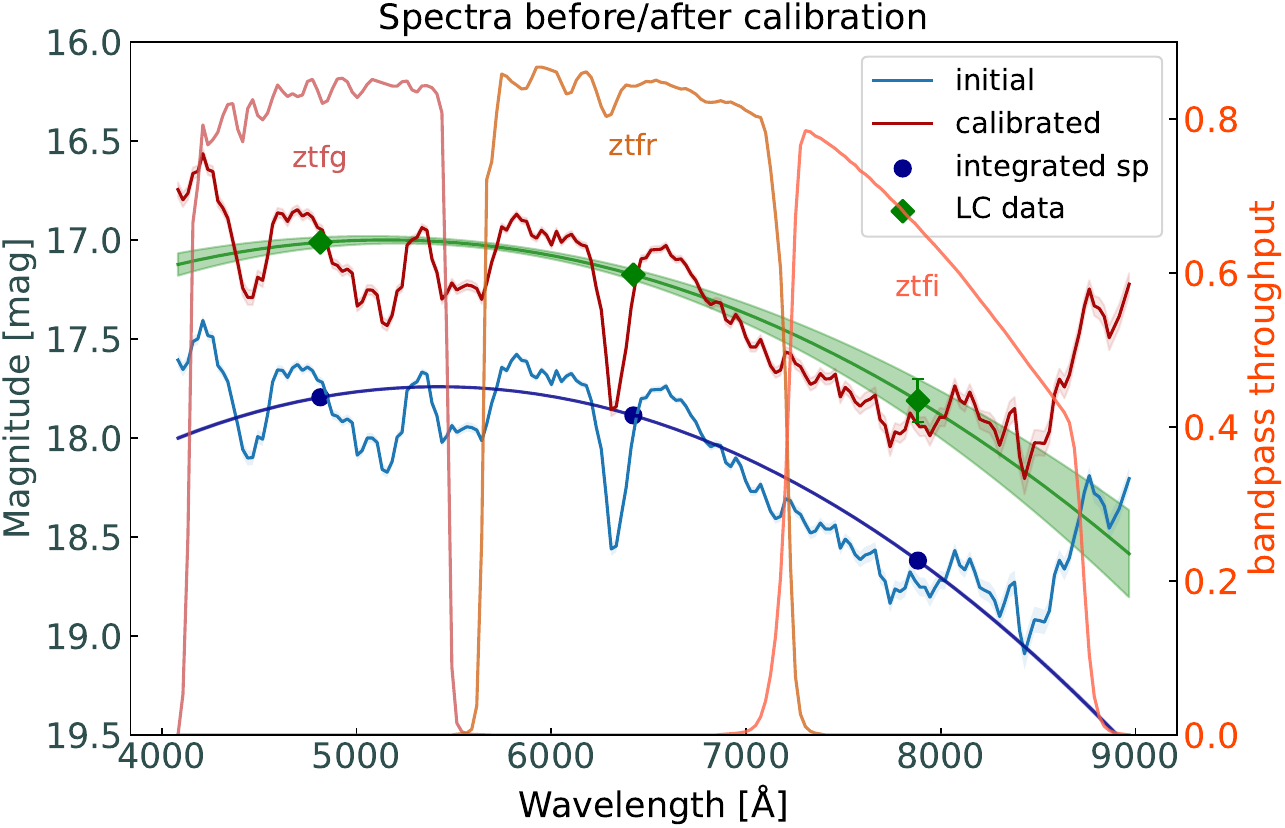}
    \caption{Spectrum of SN \textit{ZTF20aayvubx} at phase $0.01$ day (in blue) and after flux calibration (in red). The camera filter bandpasses are displayed in orange. The blue dots show the spectrum integrated through the filters. The green dots are the LC values. The offsets are 0.780, 0.708, and 0.805~mag in $g$, $r,$ and $i,$ respectively. The second-order polynomials that pass through the three photometric points are shown in their respective colors.}
    \label{fig:calibration_process}
\end{figure}

We used the calibrated LC data to further calibrate the spectra. For each SN spectrum at phase $p$, we had access to the flux values of the three interpolated LCs at the corresponding time of observation. We compared the synthetic photometry, computed with \texttt{sncosmo} \citep{barbary2016a} in the ZTF camera filter bandpasses to these three interpolated fluxes, as illustrated in Fig.~\ref{fig:calibration_process}. 

\begin{figure}
\centering
    \includegraphics[width=19pc]{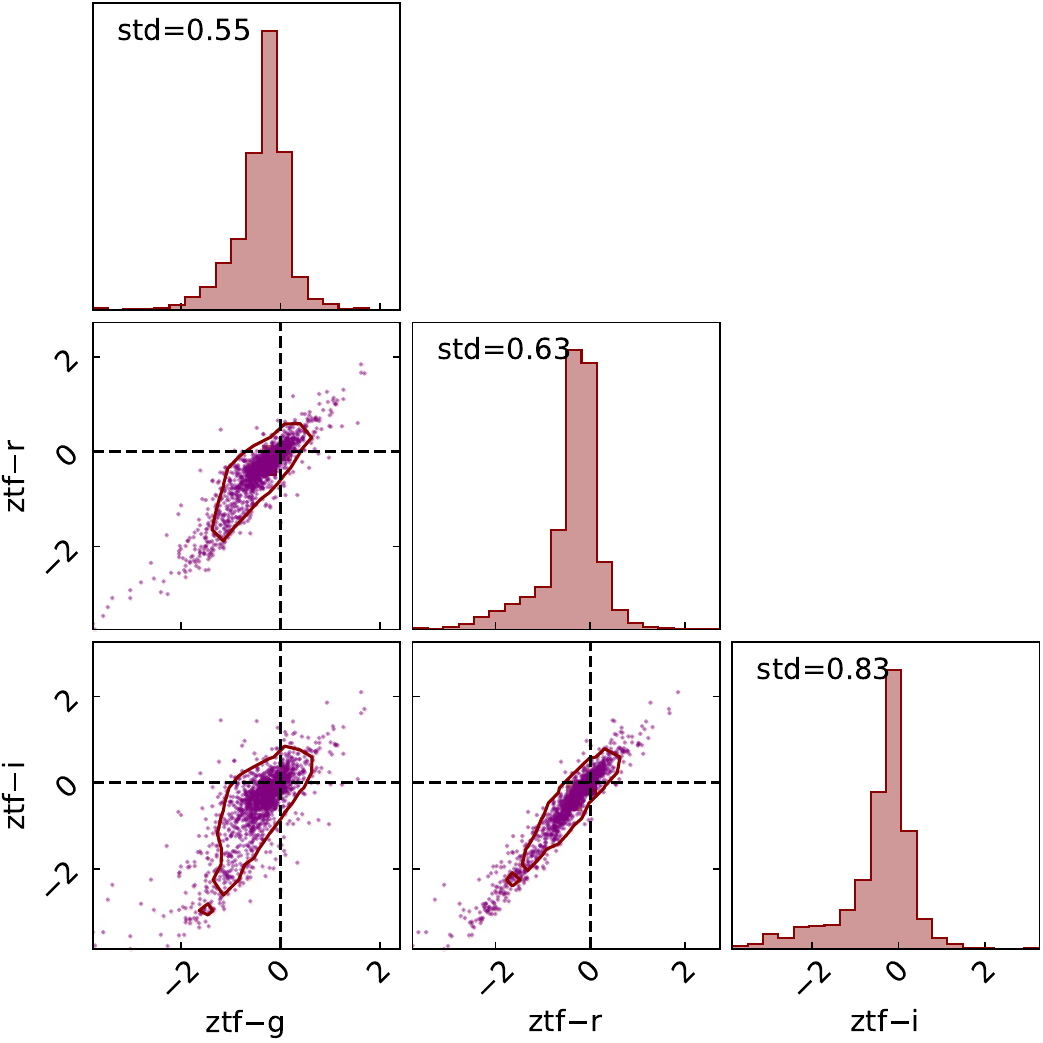}
    \caption{Magnitude offsets between the 1897 ZTF integrated spectra and the associated LC values before calibration in $g$, $r$, and $i$ filters. The STD are shown in the legend. The dashed black lines indicate the LC expectations, and the 2$\sigma$ density levels of the 2D histograms are shown in dark red.}
    \label{fig:cornerplots}
\end{figure}

 We obtained three magnitude offsets from the $gri$ observations through which we passed a polynomial of degree two and corrected the spectrum for this polynomial. We took the synthetic photometry error and LC data-point errors into account during the fit. The offset distributions before calibration are shown in Fig.~\ref{fig:cornerplots} in magnitude. Their dispersion are about $\pm0.55$, $\pm0.63$, and $\pm0.83$~mag on average in $g$, $r$, and $i$, respectively. We noted a strong correlation that propagates as an error on the polynomial of the calibration. We used a degree two, as it is the maximum degree for fitting three data points, including for SNe without $i$ data, because the offset uncertainty in $i$ was taken into account in the filter interpolation.

 For numerical stability, we used orthogonal Legendre polynomials and fit the coefficients on the spectra in magnitude. We show the coefficient distributions in the upper panel of Fig.~\ref{fig:dlr_coefs}, as well as their mean and standard deviation (hereafter STD) values. Here, $a_0$ corresponds to a global magnitude offset, $a_1$ to a slope, and $a_2$ to a quadratic correction.

\subsubsection{Precision of the flux calibration}
\label{subsec:calib_result}

To evaluate the spectrophotometric precision of this procedure, we conducted two studies. First, we provided an estimate of the SEDm observation errors alone, including telluric line residuals and instrumental effects. A second study included SN observation errors, which are mainly residuals from the host galaxy spectrum. 

\paragraph{Estimation based on standard stars}
\label{subsec:calib_result_star}

 The SEDm data acquisition includes observations of standard stars at a daily frequency to derive a primary flux calibration. Their fluxes are known and considered constant in time. We applied the calibration procedure on all the spectra of each of six standard stars (Feige34, G191B2B, HZ2, HZ4, HZ44, and LB227) for a total of 518 spectra based on their broadband magnitudes computed from the reference catalog of CALSPEC standard star spectra \citep{bohlin2020a}. Then, we compared the newly flux-calibrated spectra to the reference spectra.
 
\begin{figure}
\centering
  \includegraphics[width=\linewidth]{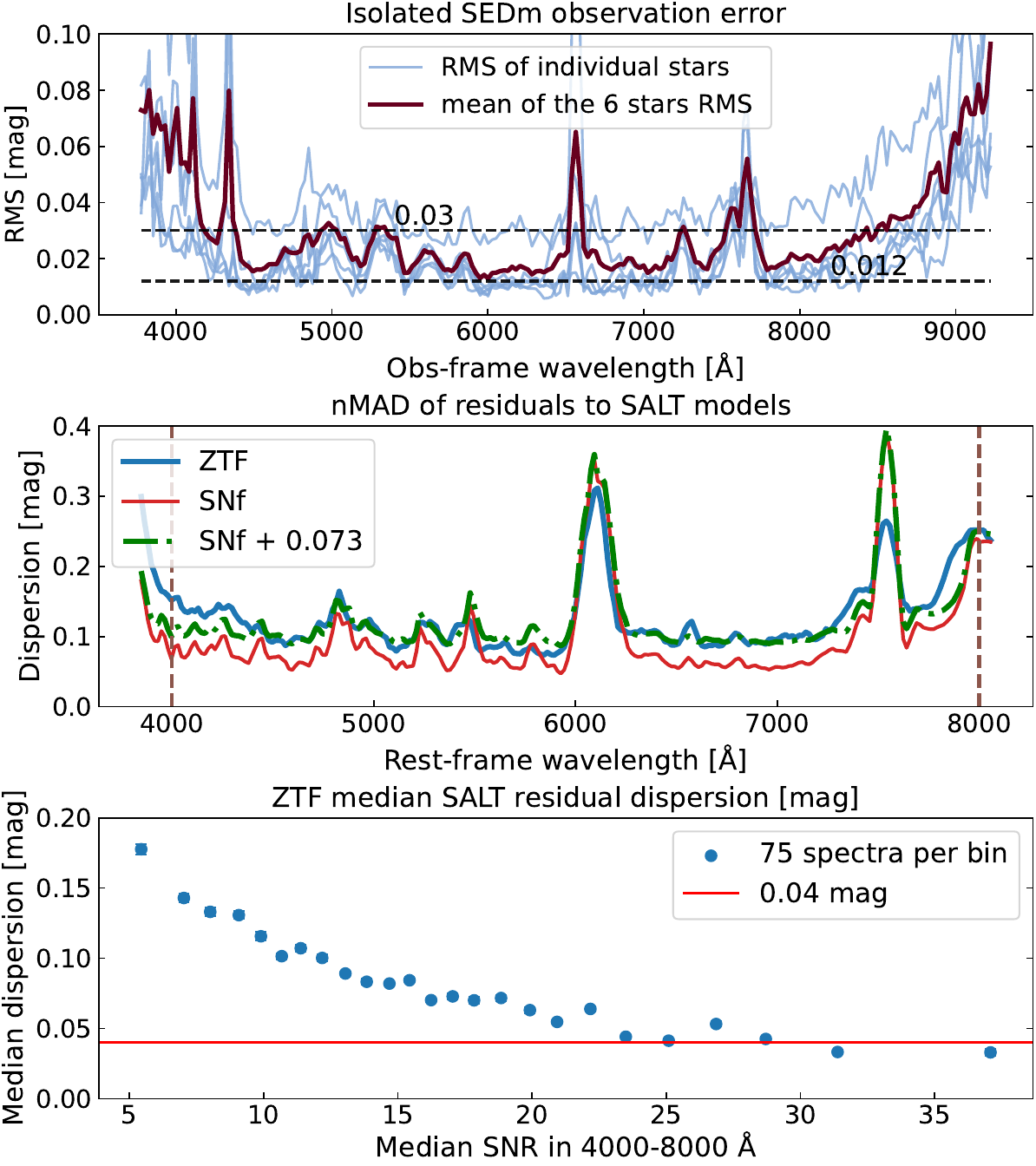} 
  \caption{Top: Precision (RMS) of the flux calibration as a function of wavelength for six standard stars (light purple) and the quadratic average (red). The limits of 0.03~mag and 0.012~mag are shown as dashed lines. Middle: Dispersion (nMAD) of the residuals relative to a SALT 2.4 spectrum model given LC data for 1897 ZTF spectra and 583 SNfactory spectra. The nMAD of SNfactory (light green) plus an additional 0.07~mag noise from ZTF are also shown. The wavelengths delimiting a good precision are shown with dashed lines at 4000~\AA{} and 8000~\AA{}. Bottom: Median SALT model residual dispersion (nMAD) binned in SNR for the ZTF (75 SNe per bin). A floor of 0.04~mag is shown by the red line. }
  \label{fig:calib_salt_model_sp}
\end{figure}

 The dispersion of the residuals of the star calibrated spectra with their reference spectra, shown in the upper panel of Fig.~\ref{fig:calib_salt_model_sp}, gives an estimate of the calibration precision in flux. We computed it for each star and then averaged it out. We obtained a spectral RMS between 0.012~mag and 0.03~mag in the 4000--8500~\AA{} range. 
 The dispersion is larger for lower and higher wavelengths because of the instrumental response and quadratic flux calibration, illustrated in Fig.~\ref{fig:calibration_process}, especially at large wavelengths because the photometry in $i$ is less precise. Atmospheric absorption lines are visible: the double peak of the dispersion around 7600~\AA{} corresponds to the Fraunhofer A $\text{O}_2$ atmosphere absorption line, and the smaller peak next to it is the $\text{H}_2\text{O}$ absorption line at 7200--7300~\AA{}. A varying absorption line of the stars is visible as well: the $\text{H}\alpha$ feature at 6563~\AA{}. 
 
This process allowed us to derive a lower limit of the calibration precision because it does not contain the host subtraction errors and because SNe are fainter than our standard stars; the latter are detected at $\sim$11 to 15~mag in Bessell-$V$ and at a mean of $\sim$18~mag for Type Ia SNe. Nevertheless, it is a good approximation for a bright isolated SN.

\paragraph{Estimation based on SNe} 
\label{subsec:calib_result_snia}

SALT2.4 \citep{guy2007, betoule2014} provides SN~Ia spectral models for given LC data, a spectrum phase, and a wavelength range. Using the \texttt{sncosmo} library \citep{barbary2016a}, we computed a spectral model for a given SN~Ia spectrum by providing its associated SALT2 parameters derived from LC data. 
The method for isolating the ZTF spectrum flux-calibration error is the following. We compared the ZTF SNe~Ia spectra to their associated SALT spectrum model and computed the dispersion of their residuals. This dispersion in wavelength originates from the SN spectrum error and the SALT model error. To isolate SN spectrum errors for ZTF, we subtracted the SALT model error. A proxy for this error was derived by computing the dispersion of the residuals of the SNfactory spectra with their SALT models, under the hypothesis that their flux calibration error is negligible. 

We computed the dispersion of the spectrum model residuals for the ZTF sample using only 1897 spectra from 1607 SNe~Ia after the cuts of the DR2 \citep{rigault2025} that are described in Table~\ref{tab:table_cut}. The LC data we provided to \texttt{sncosmo} are described in \cite{smith2025}. This dispersion is shown in Fig.~\ref{fig:calib_salt_model_sp}. We also computed the SALT residuals for the 583 SNfactory spectra from 203 SNe used in \citetalias{boone2021}. The residual dispersion shows a similar response in wavelength to that of ZTF, as shown in Fig.~\ref{fig:calib_salt_model_sp}, because the SALT model is limited at the location of strong SN lines.

 To estimate the full ZTF spectrum precision after calibration, we added an error floor in quadrature to the SNfactory SALT residual dispersion. A floor of $0.07$~mag causes the dispersions to match at most wavelengths, between 4000 to 8000~\AA{}, as illustrated in Fig.~\ref{fig:calib_salt_model_sp}. As ZTF is flux calibrated using SALT parameters, its residuals are closer to zero by construction than those of SNfactory, and we therefore focused on residual scatter (nMAD) and not on the total RMS. After the calibration process, the remaining ZTF error was almost achromatic, at 0.07~mag on average. 

 There are two contributions to this additional scatter at 0.07~mag from ZTF compared to SNfactory: a flux-calibration error floor, and an error scatter due to the weak SNR of ZTF spectra ($\sim~18$ per wavelength bin). To separate these contributions, we reproduced the aforementioned analyses per bin of SNR, and we show the evolution of the scatter in SNR in the lower panel of Fig.~\ref{fig:calib_salt_model_sp}. We computed the average (median) nMAD per SNR bin and on a wider wavelength range of [4000, 8000]~\AA{} because the median ignores the strong variations from SN features. As expected, the magnitude scatter decreased as the SNR increased, until it reached a floor of $0.04~\mathrm{mag}$ at $\mathrm{SNR}~\sim~25$. This floor corresponds to the precision in flux calibration with respect to SNfactory, and it is consistent with the floor estimated from isolated standard stars (see Sect.~\ref{subsec:calib_result_star}). This calibration error adds up quadratically to that of the SNR (therefore 0.057~mag), to explain the total ZTF spectral error after calibration, giving a total of 0.07~mag on average.
 
\subsubsection{Impact of host contamination}
\label{subsec:impact_host}

To estimate the amount of host galaxy contamination in the total spectrum extraction error, we studied the correlation between the position of the SN in the galaxy and the calibration coefficients, introduced in Sect.~\ref{subsec:calib_method}. The host subtraction is more critical in the inner parts of the galaxy. We used the $d_{DLR}$ \citep{sullivan2006, gupta2016a}, which is defined here as the normalized distance to the galaxy center, where zero is the center, and one is the luminosity radius of the galaxy in the direction of the SN. 

\begin{figure}
    \centering
    \includegraphics[width=\linewidth]{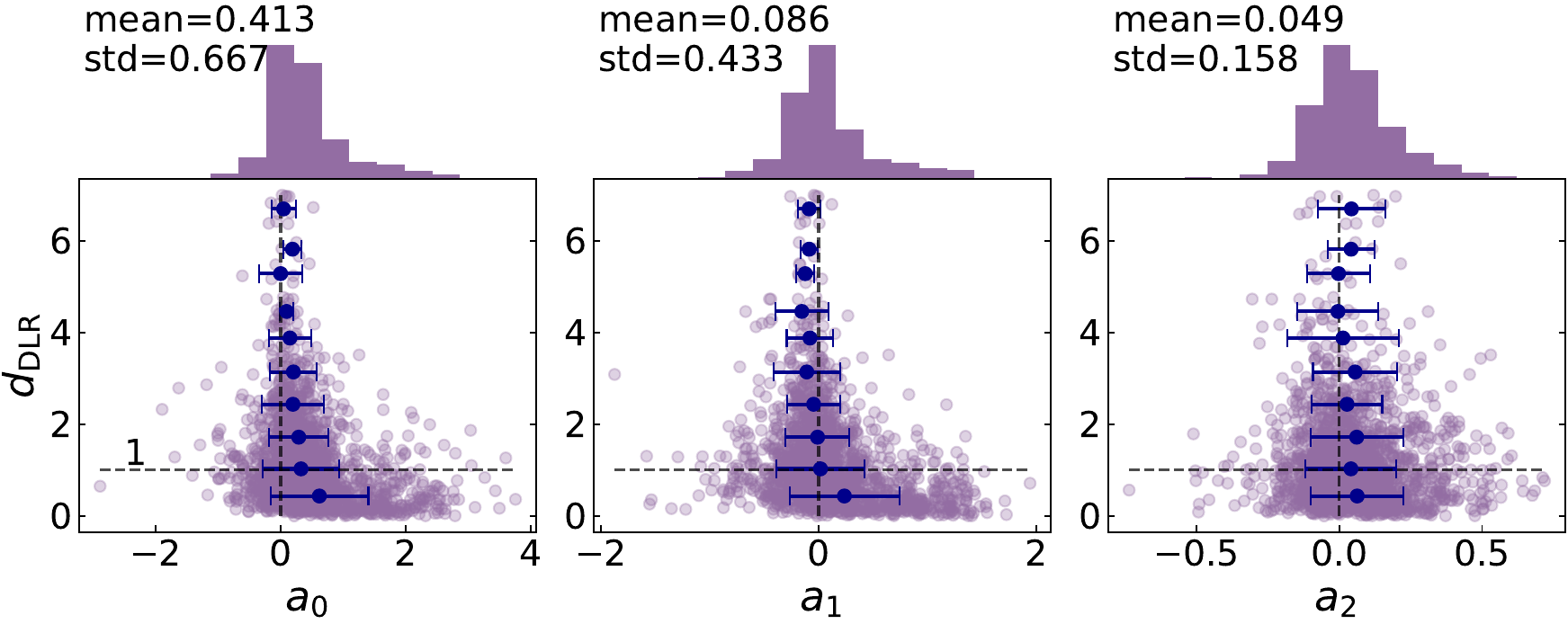}
    \caption{Top panels: Distributions of the calibration coefficient $a_0$, $a_1,$ and $a_2$ for 1897 spectra. Their mean and STD are indicated in the legends. Bottom panels: $d_{DLR}$ against the calibration coefficients. The dark blue data points show the coefficient mean and STD binned on 10 $d_{DLR}$ values. The vertical dashed black lines indicate the values corresponding to no correction, and the horizontal lines correspond to the $d_{\text{DLR}}=1$ limit. }
    \label{fig:dlr_coefs}
\end{figure}

  The coefficients as a function of the $d_{DLR}$ values are shown in Fig.~\ref{fig:dlr_coefs} for 1897 spectra of 1607 SNe after DR2 cuts. We note that the dispersion of the $a_0$ and $a_1$ coefficients increases for lower $d_{DLR}$, and in particular, below 1. As the dispersion spreads, the mean value is also shifted to higher values. This means that the closer the SN to the host galaxy, the more significant the calibration correction. The quadratic coefficient $a_2$ is poorly correlated with the $d_{DLR}$, as its Pearson coefficient is $r=-0.073\pm0.023$, while $r=-0.238\pm0.022$ for $a_0$ and $r=-0.262\pm0.022$ for $a_1$.

A significant part of the calibration correction, and therefore, of its error, is due to the host galaxy residuals, especially as a magnitude offset and a slope. There is a correction bias, correlated with $d_{DLR}$, and SNe with low $d_{DLR}$ parameter are thus more likely to have a large flux error.

\subsection{The spectrophotometric ZTF-SEDm SN~Ia sample}
\label{subsec:pres_sample}

We provide a selected sample of spectra after flux calibration. It is an homogeneous sample from the DR2 spectra, only observed with the SEDm. It consists of 1897 spectra from 1607 SNe~Ia. We removed the poorly calibrated spectra with a cut on the calibration coefficients, as detailed in Table~\ref{tab:table_cut}. We propagated the polynomial variance to the calibrated spectra, neglecting non-diagonal covariance terms. Similarly to \citetalias{boone2021}, we applied a cut on SNR defined in the bluest and reddest values of the spectrum, which we label $\text{SNR}_{\text{poly}}$, to take this polynomial error into account. We defined it in the 1000~\AA{} extrema values, and 7\% of the sample were removed as a result. This error was clearly separated from the spectrum error in order to properly maintain the spectrum SNR.

\begin{figure}
    \centering
    \includegraphics[width=17pc]{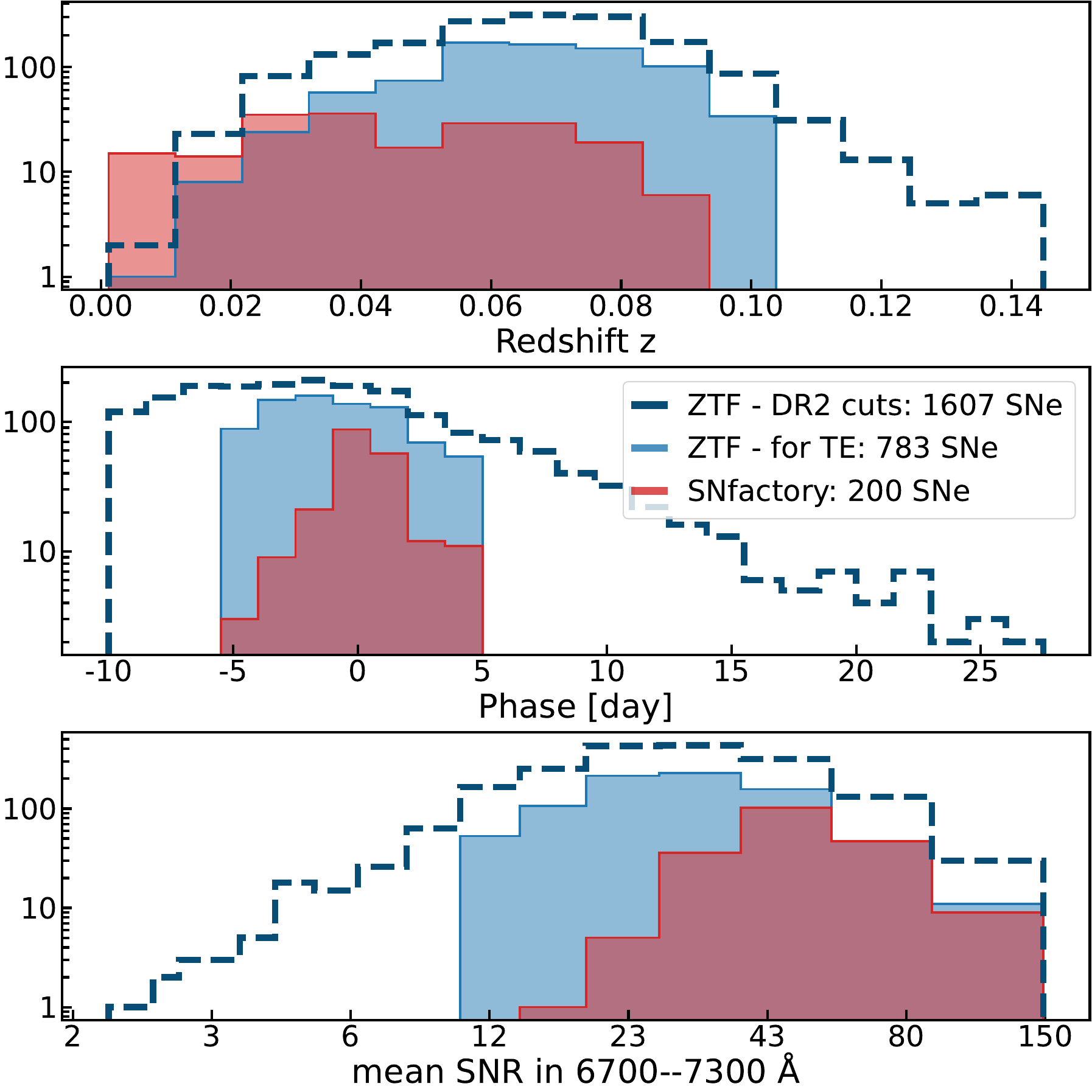}
    \caption{Distributions of SN redshifts, phases nearest maximum, and SNR defined in the range 6700--7300~\AA{} for the flux-calibrated sample of ZTF SNe~Ia after DR2 cuts (dashed blue line), the subsample for TE application (blue), and the SNfactory sample (red). }
    \label{fig:ztf_snf_distrib}
\end{figure}

The spectroscopic sample was flux calibrated using photometric data whose accuracy is at the percent level, and the final average dispersion was estimated to be 0.07~mag in the 4000--8000~\AA{} range. This combined a common error floor of 0.04~mag and a contribution from the lowest SNRs. The redshift, phase, and SNR distributions of this sample are shown in Fig.~\ref{fig:ztf_snf_distrib}. We note that the redshift accuracy varies from $\sim~10^{-5}$ to $\sim~10^{-3}$ depending on its source. At this stage, spectra are still in observer frame and are not corrected for Milky Way extinction. The flux-calibrated sample is available online~\footnote{https://ztfcosmo.in2p3.fr/}.

\section{The spectroscopic standardization method}
\label{sec:twins_embedding_method}

This section summarizes the TE spectrophotometric standardization method \citepalias{boone2021} to introduce the concepts needed when it is applied to the ZTF SEDm data. Similarly to the photometric SN~Ia standardization, the first step of the TE spectroscopic standardization is to estimate parameters that explain the input SN~Ia data. These are the stretch $x_1$ and color $c$ in the SALT-based photometric analysis \citep{guy2007, guy2010}, while these are three $\xi_{1,2,3}$, and a color (named $\Delta A_V$) for TE, as detailed in \citetalias{boone2021}. One advantage of the TE method is the access to the rest frame to derive its parameter. While the TE method aims to standardize SNe~Ia using spectra alone, it uses the phase, which is derived from photometric data for now. 

The second standardization step infers correlations between the $\xi_{1,2,3}$ and $\Delta A_V$ parameters and the SN~Ia peak brightness, thereby reducing the scatter along the Hubble diagram by empirically accounting for observed variabilities. We summarize these two steps in the next subsections.

\subsection{Spectroscopic parameters \texorpdfstring{$\Delta A_V$}{A_V} and \texorpdfstring{$\xi$}{E} }
\label{sec:te_parameter_estimation.}

The TE parameterization consists of three steps: first, to correct for spectral variations in phase to obtain spectra at maximum light, second, to fit a RBTL color term $\Delta A_V$, and third, to extract nonlinear Isomap parameters from spectral features. In principle, steps 2 and 3 might be made simultaneously, but this was not the approach chosen by \citetalias{boone2021}. As discussed in Sect.~\ref{sec:application_of_TE_on_ZTF}, no TE vector was retrained for this analysis; we used them as given by \citetalias{boone2021}.

\subsubsection{Phase correction} 
\label{sub:te_phase_correction}

The objective of this first step is to estimate the spectrum at maximum light per SN~Ia. \citetalias{boone2021} approximated the quadratic evolution in phase of SN~Ia per wavelength with respect to maximum light within the $\pm5$ days rest-frame phase range,
\begin{equation}
\begin{aligned}
    m_i(\lambda,p) - m_i(\lambda, 0) &= c_1(\lambda) \cdot p + c_2(\lambda) \cdot p^2 ,
\end{aligned}
\label{eq:max_light_formula}
\end{equation}
with $m_i(\lambda, p)$ the spectrum of SN $i$ expressed in magnitude at rest-frame phase $p$ as a function of rest-frame wavelength $\lambda$; and $c_{1,2}(\lambda)$ two spectroscopic model coefficients to correct for the phase $p$. This differential time evolution model (hereafter DTEM) only corrects for variability that is common to all SNe~Ia and not for variability that is specific to individual targets. \citetalias{boone2021} reported that for the SNfactory dataset, and averaged over all wavelengths,  84.6\% of the variance in the evolution of spectra near maximum light is common to all SNe~Ia and is thus captured by the DTEM. The DTEM is independent of photometric LC data; nevertheless \citetalias{boone2021} also proposed a DTEM with additional stretch corrections. We retained the standard model for our baseline. The extended DTEM, which takes into account the stretch, is analyzed in Appendix~\ref{app:DTEM_with_stretch}.

Finally, when several spectra for a given SN~Ia were available within this $\pm5$-day phase range, they were combined after phase correction to have only one spectrum at maximum light per target. This takes the error of the model into account, which is modeled as a function of the phase (details in \citetalias{boone2021}). Once the vectors of this model have been trained on SNfactory, the DTEM can simply be applied to other samples, so we used those from \citetalias{boone2021}.

\subsubsection{Read between the lines}
\label{subsec:te_rbtl}

After we had one spectrum per SN~Ia, corresponding to the spectrum expected at maximum light, we estimated the color component $\Delta A_V$. We used the method called read between the lines introduced by \citetalias{boone2021}. 

This method captures the color term ignoring the strong absorption lines, which exhibit great dispersion from one SN~Ia to the next. Outside of these well-identified lines at rest frame, the spectrum of SN $i$ at maximum light ($f_{\text{max},i}(\lambda)$) can be modeled as $f_{\text{model},i}(\lambda)$, 
\begin{equation}
    -2.5 \log_{10} \left(\frac{f_{\text{model}, i}(\lambda)} { f_{\text{ref}}(\lambda)} \right) = \Delta m_i +  \Delta A_{V, i} \cdot CL(\lambda) ,
    \label{eq:rbtl_model_1}
\end{equation}
with a reference SN~Ia spectrum at maximum ($f_\text{ref}(\lambda)$), reddened by an established dust-color law $CL(\lambda)$ with the coefficient $\Delta A_{V,i}$, and with a gray offset $\Delta m_i$. 

In practice, $f_{\text{ref}}(\lambda)$ was assumed to be the average SN~Ia spectrum at maximum of light of the SNfactory sample. The strong spectral features with large dispersion, common to all SNe~Ia, were identified by \citetalias{boone2021} by estimating a wavelength-dependent intrinsic scatter $\eta(\lambda)$, defined as a fraction of flux, or a difference in magnitude. Wavelengths at which this scatter was strongest were de-weighted during the fit. After $\eta(\lambda)$ has been modeled, strongly varying wavelengths can be fully discarded with a Boolean cut on $\eta$. This was not considered in \citetalias{boone2021}, but we illustrate the benefits of this stricter RBTL approach in Sect.~\ref{sec:RBTL}.

The RBTL parameters $\Delta A_{V,i}$ and $\Delta m_i$ were fitted for each SN~Ia with a simple $\chi^2$ minimizing $f_{\text{max},i}(\lambda)$ and $f_{\text{model},i}(\lambda)$ given measurement errors (see details in Sect.~4.1 of \citetalias{boone2021}). The de-reddened SN~Ia spectrum is thus defined  as
\begin{equation}
  f_{\text{dered},i}(\lambda) = f_{\text{max},i} (\lambda) \cdot 10^{+0.4(\Delta m_i + \Delta {A}_{V,i} \cdot CL(\lambda)) }.
  \label{eq:rbtl_dered}
\end{equation}

\subsubsection{Manifold learning}
\label{sec:te_manifold}

The last step of the TE parameterization estimates the $\xi$ parameters that aim to describe variabilities in SN~Ia spectra at maximum, not already accounting for the RBTL parameters. Following the initial concept of twin spectra \citep{fakhouri2015}, the TE is constructed based on spectral distances between SNe~Ia that are input to an Isomap algorithm \citep{tenenbaum2000}. This algorithm computes a nonlinear low-dimensional reduction known as manifold learning, which was shown to outperform a classical (linear) PCA in its ability to capture data variance in a smaller number of eigenvectors. 

For the SNfactory data, the fraction of explained variance reaches a limit at the third component (see Sect.~5.1 of \citetalias{boone2021}). With three components $\vec{\xi}$, this manifold technique can account for 86.6\% of the intrinsic SN~Ia dispersion. This explained dispersion is mostly contained within strong and highly variable spectral regions of SN~Ia spectra, reducing the scatter on \ion{Ca}{II} and \ion{Si}{II} features from $\sim$0.3~mag to less than 0.1~mag (see \citetalias{boone2021}~Fig.~6), for example.

In practice, when the Isomap was trained on SNfactory data, the process of estimating the parameters $\vec{\xi}$ for another SN~Ia sample was the following: the spectrum $f_{\text{dered},i}(\lambda)$, corrected for RBTL color and magnitude, was input to the trained Isomap algorithm from the \texttt{sklearn} library~\citep{pedregosa2011}, and three parameters $\vec{\xi}_{1,2,3}$ were returned per SN.  

\subsubsection{Modeling of spectral variation}
\label{subsec:GP_spectral_method}

An extension of the spectral model described in Eq.~\ref{eq:rbtl_model_1}, making use of the $\vec{\xi}_i$ spectral impact, was developed by training a Gaussian process (hereafter GP). It returns a single value per wavelength as a ratio of the flux for given Isomap coordinates. This allowed us to model the spectrum of SN $i$ at maximum light more accurately, and Eq.~\ref{eq:rbtl_dered} becomes
\begin{equation}
  f_{\text{model},i}(\lambda) = f_{\text{ref}} (\lambda) \cdot 10^{-0.4(\Delta m_i + \Delta {A}_{V,i} \cdot CL(\lambda) )} \cdot (1+\text{GP}(\lambda)(\vec{\xi}_i)).
  \label{eq:TE_GP_formula}
\end{equation}

This model was not developed as a standardization method in \citetalias{boone2021a} because it does not intend to decrease the overall gray dispersion. Instead of modeling the spectral variation through this GP per wavelength, \citetalias{boone2021a} therefore used an alternative method to explain the gray offset, as presented in the next subsection.

\subsection{Twins-embedding standardization}
\label{sec:te_standardization.}

After we defined the RBTL (Sect.~\ref{subsec:te_rbtl}) and manifold learning (Sect.~\ref{sec:te_manifold}), we introduce the TE standardization method from \citetalias{boone2021a}. This spectroscopic standardization relies on the Isomap parameters $\vec{\xi}$ from Sect.~\ref{sec:te_manifold}, and is implemented using a GP \citepalias{boone2021a} as
\begin{align}
    \delta m_i^{\mathrm{GP}} =  GP \left(\mu ( m_{\mathrm{ref}} , \omega \cdot\Delta A_{V, i} ) ; \sigma ( \vec{\xi}_i, \vec{\sigma}^2_{p.v., i} , \sigma^2_u )\right).
   \label{eq:residual_formula}
\end{align}
For a given SN $i$, the mean function of the GP uses a $\Delta A_V$ correction ($\omega \cdot\Delta A_{V, i}$)  alongside a gray magnitude offset $m_{\rm ref}$; the variance function of the GP uses the Isomap parameters $\vec{\xi}_i$ and the peculiar velocity $\sigma^2_{p.v., i}$, along with GP hyperparameters common to the set of SNe, such as the intrinsic dispersion $\sigma_u$. This model yielded an unexplained dispersion of 0.073~mag ($\sigma_{u}$) from the SNfactory sample of 134 SNe that met the standardization requirement cuts (detailed in \citetalias{boone2021a}). In the following study, we fixed $\omega = 0$, in contrast to \citetalias{boone2021a}, as we wished to correct for any additional color variation in a separate step. 

\section{ZTF sample for Twins Embedding} 
\label{sec:ztf_vs_snf}

The TE method was developed and tested using 173 spectrophotometric SN~Ia spectra with a high SNR from the SNfactory sample \citep{aldering2002}, rebinned to a $R\sim300$ resolution (initially $R~\sim~2000$). The method was trained by \citetalias{boone2021} on half of the sample, and it was tested on the other half at the un-blinding of the results to validate the method. It has not been applied to another dataset so far.

In this section, we apply relevant cuts to the ZTF sample for TE application and compare the resulting subsample with that from SNfactory. We ran the preliminary data processing common to SNfactory, whose full framework can be found in \citetalias{boone2021}. Finally, in this section, we interpolate the spectra to maximum light using the DTEM.

\subsection{Cuts and processing of the sample}
\label{subsec:cuts_processing_ztf}

Table~\ref{tab:table_cut} details the cuts we applied to the sample, for the full flux-calibrated sample and for the sample on which we evaluated its precision. We created two other samples: a sample of 783 SNe for the application of the TE method, and a sample of 688 SNe for the standardization using the TE. 

\begin{table}
\caption[]{\label{tab:table_cut}Cuts applied to different ZTF samples: the selection, the removed fraction, and the number of SNe remaining in the sample.}

\begin{tabular}{lccc}
        \hline \hline
        Cut & Interval & \% removed  & \# SNe \\ 
        
        \hline
        \noalign{\smallskip}
       \multicolumn{4}{c}{ Flux-calibrated sample} \\
       \noalign{\smallskip}
       \hline
        source & only SEDm & 40\%  & 1872\\
        flux coefs & \begin{tabular}{@{}c@{}}$|a_0|<4$  \\ $|a_1|<2$\\$|a_2|<0.75$\end{tabular} & 1 \%  & 1853\\
 calib. error &$\text{SNR}_{\text{poly}}>10$&7\%&1721\\
       \hline
       \noalign{\smallskip}
       \multicolumn{4}{c}{ Evaluation of the precision} \\
       \noalign{\smallskip}
       \hline
       ZTF DR2 & \begin{tabular}{@{}c@{}} $-0.2\le c\le+0.8$
        \\ $-3 \le x_1 \le +3$
        \\ $x_{1, \text{err}} \le 1$
        \\ $c_{\text{err}}  \le 0.1$
        \\ $t_{0, \text{err}} \le 1$
        \\ fitprob $> 10^{-7}$
        \end{tabular} &  7\% & 1607\\
       \hline
       \noalign{\smallskip}
       \multicolumn{4}{c}{ RBTL sample} \\
       \noalign{\smallskip}
       \hline
         phase  & $\pm 5$ days & 41\% & 939\\
       redshift & z $<$ 0.1 &  5\%  & 888\\
       peculiar & cosmo &  8\%  & 824\\
       Quality & SNR $>$ 10 & 3\%  & 783\\
       \hline
       \noalign{\smallskip}
       \multicolumn{4}{c}{ standardization} \\
       \noalign{\smallskip}
       \hline
         redshift & z $>$ 0.03 & 3\%  & 744\\
        RBTL &
       \begin{tabular}{@{}c@{}} 
       $-0.5 <\Delta A_V< 0.5$ 
      \end{tabular} & 9\%  & 688 \\ 
       \hline
    \end{tabular}

\end{table}     

For the TE sample of 783 SNe, we kept good-quality data, that is, data with a cut on redshift above 0.1 and a cut on $\text{SNR}<10$ that we defined in the wavelength range of 6700--7300~\AA{} as the mean of the ratio of the flux and the STD of the fluctuations, so corresponding to an SNR per pixel. The largest cut was on the phase because the TE applies to spectra within $\pm$5 days from peak. The spectra outside this interval represent 41\% of the sample. 

Distributions of redshift, phase, and SNR are presented in Fig.~\ref{fig:ztf_snf_distrib} for the ZTF flux-calibrated sample, the TE sample, and for the SNfactory sample. The figure shows that the ZTF sample has more high-redshift SNe than the SNfactory sample. The phase distribution is more uniform for ZTF, with more SNe observed before maximum. Most SNe in the SNfactory sample have a spectrum at one day from maximum. The SNR is greater for the SNfactory sample, with a median of 49, while the median is 29 for ZTF. 

We prepared the ZTF spectra before applying the TE. We corrected for the recession speed by shifting the spectra to the rest frame. As the rest-frame wavelength range is different for each SN, we interpolated them to a common range of 4103--8388~\AA{} by keeping the same binning as \citetalias{boone2021} at 1000 km.$\text{s}^{-1}$.
The spectra were also shifted in flux assuming some cosmology to a common redshift, $z_{\mathrm{ref}} = 0.05$, in order to capture the dispersion of the magnitudes uncorrelated with redshift. At this redshift range, the cosmological model used to de-redshift does not significantly affect the result. 
 The spectra were corrected for the Milky Way extinction using the Cardelli extinction law \citep{cardelli1989} and extinction maps from \cite{schlafly2011a}, also assuming a total-to-selective extinction ratio $R_V=3.1$. 

 The SNe with a low redshift accuracy ($\sim~10^{-3}$) have an additional spectral uncertainty when the spectra are shifted to rest frame and corrected in flux to the common redshift of 0.05. It was not propagated in the spectrum variances and will need to be taken into account separately.

\subsection{Wavelength range for TE}
\label{sec:application_of_TE_on_ZTF}

As the ZTF sample is defined in the 4103--8388~\AA{} wavelength range, we truncated the TE vectors as $c_{1,2}(\lambda)$ or $f_\mathrm{ref}(\lambda)$. These were trained on the SNfactory spectra, which were defined in rest frame on the 3305--8586~\AA{} wavelength range. To compare the TE application results between ZTF and SNfactory, we computed the results in parallel for SNfactory spectra in the common wavelength range. 

By applying the TE method to the ZTF sample of 853 spectra from 783 SNe~Ia step by step, we wished to analyze the method, benefiting from the large sample size, and to determine whether the accuracy of our data is sufficient for a spectrophotometric standardization. 

\subsection{Phase correction}
\label{subsec:phase_corr}

The sample contains spectra at rest-frame phases between $-5$ and $+$5 days. To correct for spectral time evolution, we used the DTEM presented in Sect.~\ref{sec:twins_embedding_method}. It provides one spectrum at $t_{\mathrm{max}}$ per SN, using the wavelength-dependent phase evolution described in Eq.~\ref{eq:max_light_formula}. The DTEM error in phase is propagated to the spectrum uncertainty. When multiple spectra were available for one SN, the spectrum at maximum was computed by combining the corrected spectra (see details in \citetalias{boone2021}). After this procedure, the ZTF sample contained 783 spectra, one for each of the 783 SNe~Ia.

\begin{figure}
    \centering
    \includegraphics[width=\linewidth]{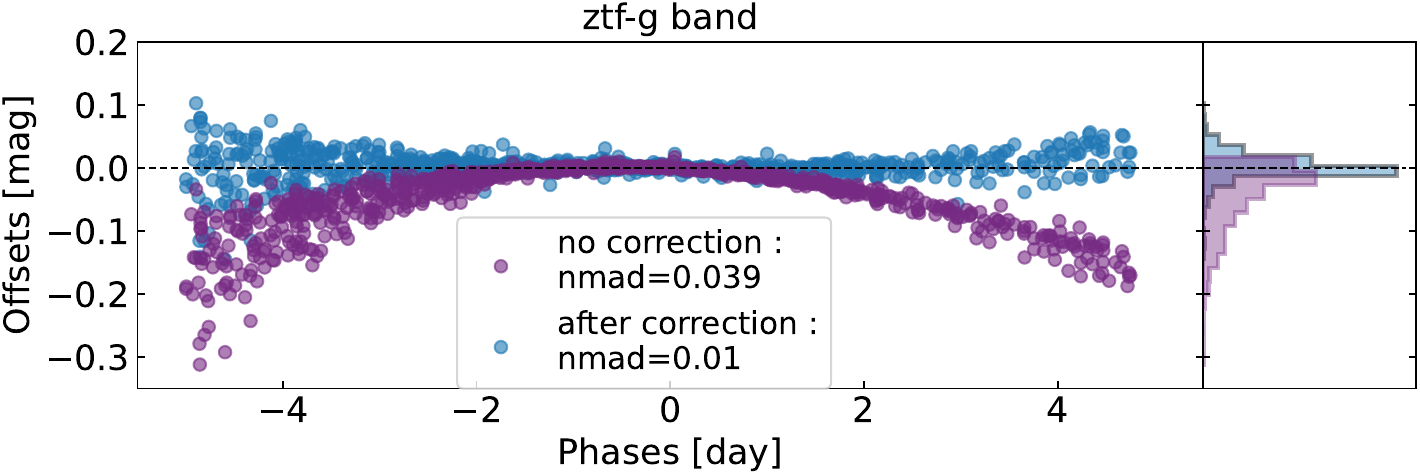}
    \caption{Magnitude offsets between the LC value at $t_{\mathrm{max}}$ in the $g$ band and integrated spectrum before (purple) and after (blue) the DTEM correction for 783 SNe as a function of the initial phase of the spectrum. The two distributions are shown in the right plot.}
    \label{fig:maxoffsets_after}
\end{figure}

To estimate the precision of the DTEM, we compare in Fig.~\ref{fig:maxoffsets_after} LCs values at maximum light, interpolated by SALT2.4, to the synthetic magnitude from spectra before and after phase correction as a function of the initial phases of the spectra. We computed it in the $g$ filter because it is closest to the Bessell-$B$ bandpass on which the maximum luminosity is defined. This allowed us to test for evolution in phase: before DTEM correction, we therefore expect the photometric flux to reach a maximum around $t_{\mathrm{max}}$. After correction with the DTEM, a spectrum integrated in this bandpass should return a constant value for any initial phase. When there were several spectra per SNe, we computed the offset for the spectrum at the phase nearest to the maximum.

Before correction, the offsets evolved with phase as expected, reaching a maximum around $t_{\mathrm{max}}$. This evolution was correctly removed by the phase-correction procedure, which displays flat residuals with phase. The dispersion (nMAD) evolves from $\sim$0.039~mag to $\sim$0.008~mag, although the model is less accurate near the phase extrema. Current use of the DTEM does not include any parameter to describe the variation in the time evolution between SNe~Ia, such as the SALT2.4 stretch parameter $x_1$, which we left out to be independent of photometry. 

By keeping only the 512 spectra (out of the 783) with a phase between $\pm 3$ days, we achieved a phase-correction residual scatter in $g$ as low as 0.007~mag.

\section{Read between the lines}
\label{sec:RBTL}

We applied the RBTL method, the concept of which is summarized in Sect.~\ref{subsec:te_rbtl}, to the 783 ZTF spectra at maximum light. We compared the results to the result obtained using the 200 spectra from SNfactory \footnote{ We cut three SNf SNe of the 203 at $z>0.1$ to compare targets at the same redshift range as ZTF.} in the same common wavelength range (see Sect.~\ref{sec:application_of_TE_on_ZTF}). We describe the method of application in Sect.~\ref{subsec:rbtl_applic_method}, and we show the results in Sect.~\ref{subsec:RBTL_results}. We show spectral residuals in Sect.~\ref{sec:rbtl_residual_scatter}, and we discuss its connection to photometry. In Sect.~\ref{subsec:RBTL_stand} we explore an initial spectrophotometric standardization method, which consists of correcting for the RBTL color parameter.

\subsection{RBTL parameter estimation }
\label{subsec:rbtl_applic_method}

We fit the two RBTL parameters (the color term $\Delta A_{V}$ and the magnitude offset $\Delta m$) for each SN $i$ independently. 

\begin{figure}
    \centering
    \includegraphics[width=\linewidth]{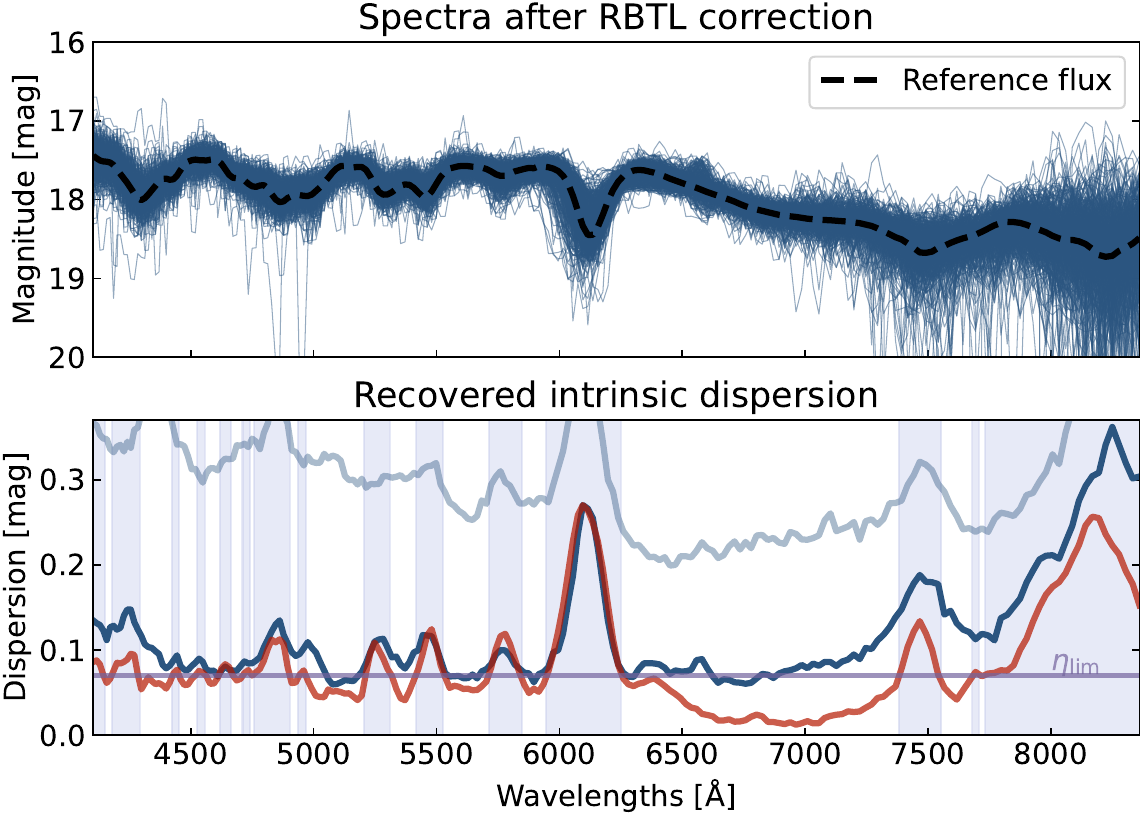}
    \caption{Spectral correction with RBTL. Top: 783 ZTF SNe~Ia spectra, corrected for RBTL parameters (dark blue), in magnitude, and the SNfactory reference spectrum (dashed black line). Bottom: Dispersion (nMAD) before correction (light blue) and after correction (dark blue). The model vector $\eta(\lambda)$ from \citetalias{boone2021}, trained on SNfactory, is shown in red. The high-dispersion areas, defined as $\eta(\lambda)>\eta_\mathrm{lim}$, and masked during the RBTL fit, are shown as vertical light purple bands in the bottom plot.}
    \label{fig:correction_rbtl}
\end{figure}

The fit was made by minimizing a $\chi^2$ for each SN. We compared the observed SN $i$ spectrum at maximum light $f_{\mathrm{max},i}(\lambda)$, and its model, which was computed using Eq.~\ref{eq:rbtl_model_1}. The reference flux is the SNfactory mean flux, $f_{\text{ref}}(\lambda)$, and is shown in Fig.~\ref{fig:correction_rbtl}. Following \citetalias{boone2021}, we used the \citetalias{Fitzpatrick1999} extinction-color relation $CL(\lambda)$ assuming $R_V=2.4$. This is close to $R_V = 2.8 \pm 0.3$ found by \cite{chotard2011}. The total error is 
\begin{equation}
 \sigma^2_{\mathrm{tot},i}(\lambda) =  \sigma^2_{\mathrm{max},i}(\lambda) + \sigma^2_{\mathrm{poly},i}(\lambda)  + \left(f_{\mathrm{model}, i}(\lambda)\cdot\eta(\lambda)\right)^2 ,
\end{equation}
with $\sigma^2_{\mathrm{max},i}(\lambda)$ the variance of the spectrum at maximum light of SN i, and $\eta(\lambda)$ the chromatic intrinsic dispersion, assumed to be a fraction of flux (i.e., close to a magnitude scatter). Both $\eta(\lambda)$ and $f_{\mathrm{ref}}(\lambda)$ were derived from the SNfactory training \citepalias{boone2021a}  (see Fig.~\ref{fig:correction_rbtl} and Sect.~\ref{sec:twins_embedding_method}). Because of the calibration process, we took an additional uncertainty into account, $\sigma^2_{\mathrm{poly},i}(\lambda)$, propagated from the calibration uncertainty (see Sect.~\ref{sec:calib}). Most of these uncertainties diverge at low and high wavelengths, as illustrated in Fig.~\ref{fig:calibration_process}, and therefore act as de-weighting the fit of RBTL parameters in these parts of the spectra.

The $\Delta m$ parameter is the gray offset left over by the color-law correction, $\Delta A_{V} \cdot CL(\lambda)$. 
For instance, $\Delta m$ receives contributions from SN~Ia common intrinsic gray scatter (estimated at $\sim0.073$~mag in \citetalias{boone2021a}), peculiar velocity scatter, and from the flux calibration precision, which we estimated for ZTF in Sect.~\ref{sec:calib}. Because the low spectral SNR affects the estimate of the overall spectral amplitude very little, we considered the contribution of the SNR to be negligible, and we only considered the spectrophotometric precision as safely contributing to $\Delta m$. The remaining variance between $f_{\mathrm{model}, i}(\lambda)$ and $f_{\mathrm{max},i}(\lambda)$ includes the SN~Ia chromatic variations (e.g., absorption line scatter), the chromatic flux-calibration error, and a potential miscorrection for the host contamination or atmospheric features. The SN~Ia absorption lines are the main sources of variance that is not explained by the RBTL model, as illustrated in Fig.~\ref{fig:correction_rbtl}.

To prevent this chromatic dispersion from being included in the SN~Ia color $\Delta A_{V}$ and the gray offset, $\Delta m$, \citetalias{boone2021} used $\eta(\lambda)$ to implement inverse variance weighting. Instead, we used it as a binary mask, passing only the wavelengths with $\eta(\lambda)<~\eta_\mathrm{lim}~=~0.073$~mag, shown in Fig.~\ref{fig:correction_rbtl}. This scatter amplitude is higher numerically for $\lambda>6500$~\AA{}. In comparison to photometry, the strength of the spectroscopic approach is the ability to remove these high SN~Ia variance regions while estimating the SN~Ia color.

\subsection{RBTL parameters} 
\label{subsec:RBTL_results}

\begin{figure}
    \centering
    \includegraphics[width=.7\linewidth]{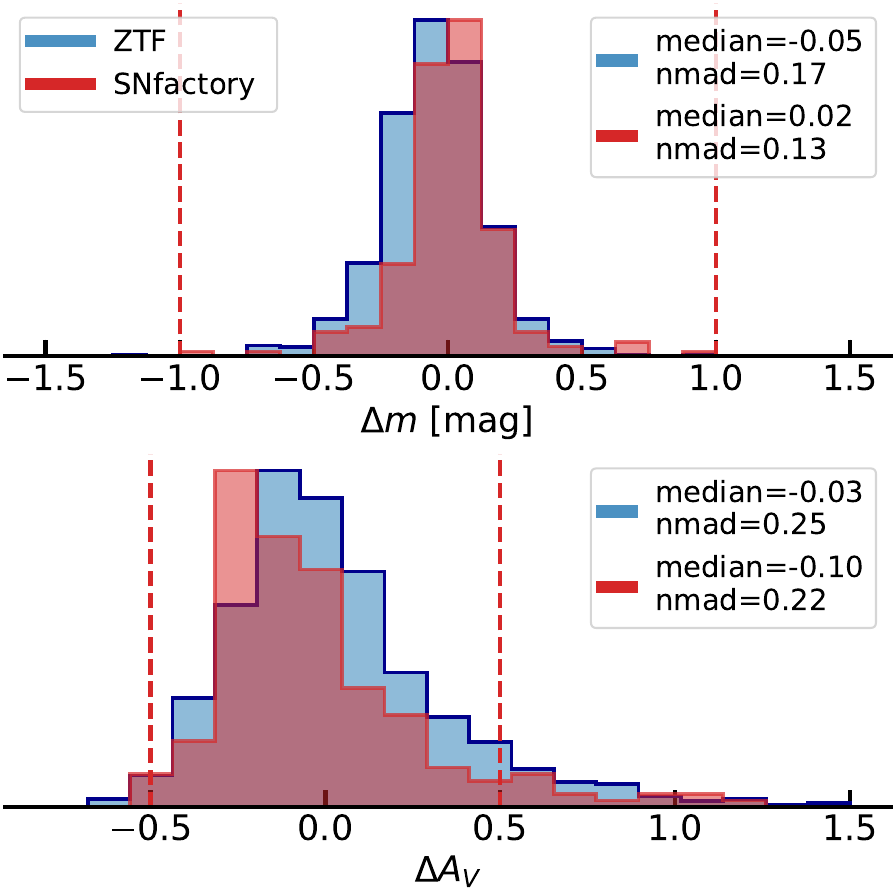}
    \caption{Normalized RBTL parameter $\Delta m$ and $\Delta A_V$ distributions for ZTF (783 SNe; blue) and SNfactory (200 SNe; red). The cuts applied for the standardization sample of those parameters alone are shown with dashed pink lines. The median and nMAD of the distributions are shown in the legends.}
    \label{fig:rbtl_with_snf}
\end{figure}

\begin{figure}
    \centering
    \includegraphics[width=\linewidth]{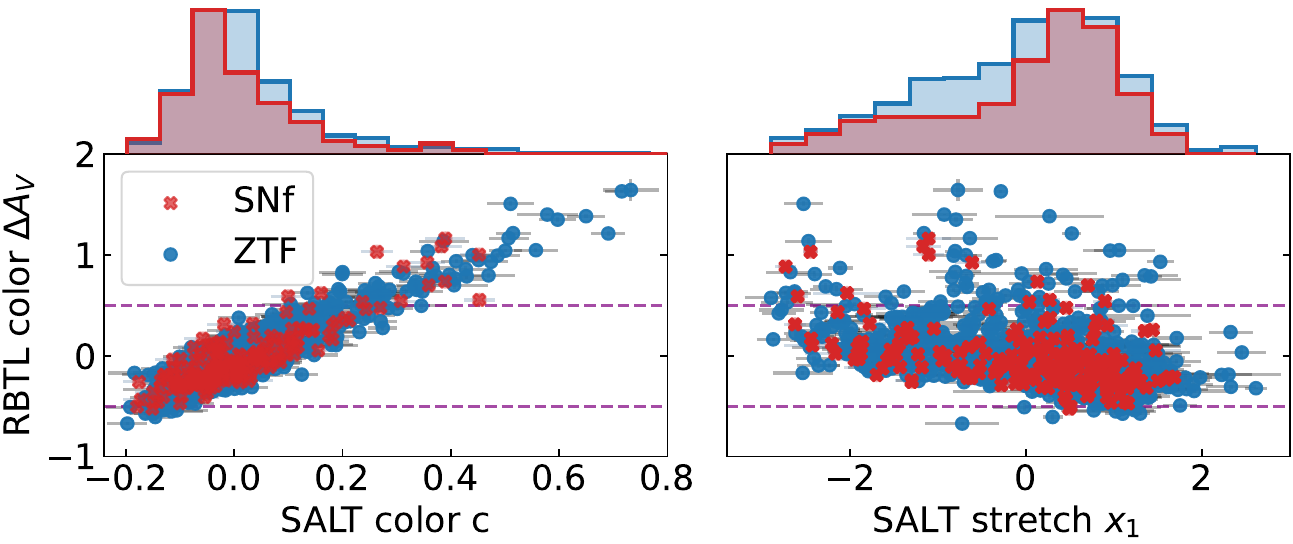}
    \caption{Read-between-the-lines color $\Delta A_V$ as a function of the SALT color $c$ and SALT stretch $x_1$ for the SNfactory sample (red) and the ZTF sample (blue). The cuts on $\Delta A_V$ for standardization at $\pm 0.5$ are shown with dashed lines.}
    \label{fig:rbtl_salt_color}
\end{figure}

We applied the RBTL method to the SNfactory sample of 200 SNe and the ZTF sample of 783 SNe. The distributions of the RBTL parameters are shown for both samples in Fig.~\ref{fig:rbtl_with_snf}. The parameter $\Delta m$ has a similar distribution, but with a narrower scatter for SNfactory (0.13~mag) than for ZTF (0.17~mag), and a shift from zero is expected for ZTF as the LCs absolute ZP is unknown (see Sect.~\ref{subsec:calib_method}). However, the $\Delta A_V$ distributions differ slightly between the two samples: ZTF contains a relatively higher number of red SNe. In comparison, the SALT2 color distributions, shown in Fig.~\ref{fig:rbtl_salt_color}, seem to be more similar in the two samples. We conducted statistical tests (e.g., $k$-sample Anderson-Darling test, hereafter A.D.) to compare the sample distributions for both colors: they show a more significant difference for the RBTL color (e.g., A.D. p-value: 0.003) than for the SALT color (e.g., A.D. p-value: 0.04). This suggests that the $\Delta A_V$ difference is also caused in part by leftover calibration issues; we discuss this further in Sect.~\ref{sec:discussion}. This might in turn explain the larger scatter on $\Delta m$ from ZTF because $\Delta m$ and $\Delta A_V$ were fitted simultaneously.

We compare the RBTL color $\Delta A_V$ with the SALT color $c$ and stretch $x_1$ parameters in Fig.~\ref{fig:rbtl_salt_color}. As expected, $\Delta A_V$, as an SN~Ia color derived from spectroscopy, is tightly correlated with $c$, which is derived from photometry. Interestingly, $\Delta A_V$ is also correlated with $x_1$: the bluest SNe are significantly slower because the stretch values are above zero, while starting at $\Delta A_V~\sim~-0.3$, the two stretch modes \citep{rigault2020} are present. This is consistent with the photometric sample results from \cite{ginolin2025a,ginolin2025}, based on the $c$ parameter, which showed that SNe in (locally) red environment are distributed equally in the two modes when the locally blue SNe are concentrated in the high-stretch mode. 

\subsection{RBTL dependence on wavelength}
\label{sec:rbtl_residual_scatter}

To estimate the spectral residuals of this RBTL fit, we applied the correction described in Eq.~\ref{eq:rbtl_dered} to both samples. We show the residual spectroscopic dispersion for ZTF and SNfactory in the bottom panel of Fig.~\ref{fig:correction_rbtl}. The dispersion corresponds to the variance that is not explained by the RBTL color+magnitude offset model. For comparison, we also show the original ZTF scatter prior to the application of the RBTL. At wavelengths bluer than 6500~\AA{}, the unexplained scatter for ZTF and SNfactory agrees remarkably well. SNfactory exhibits a net reduction in the 6500--7300~\AA{} region, as the RBTL procedure captures nearly all the SN~Ia variance at $\sim7000$~\AA{}. However, the ZTF dispersion is not reduced in this region, as it remains stable with $\sim0.1$~mag dispersion and rises to a higher dispersion at the redmost wavelengths, as does SNfactory. The ZTF spectral behavior can be reproduced using the SNfactory curve by adding a 0.057~mag error floor in quadrature (see Appendix~\ref{app:RBTL_spectral_disp}), caused by the lower SNR in ZTF spectra, as detailed in Sect.~\ref{sec:calib}. We expect that the dispersion originating from the flux calibration residuals is captured by $\Delta m$, as discussed in Sect.~\ref{subsec:rbtl_applic_method}. 

\begin{figure}
    \centering
    \includegraphics[width=0.9\linewidth]{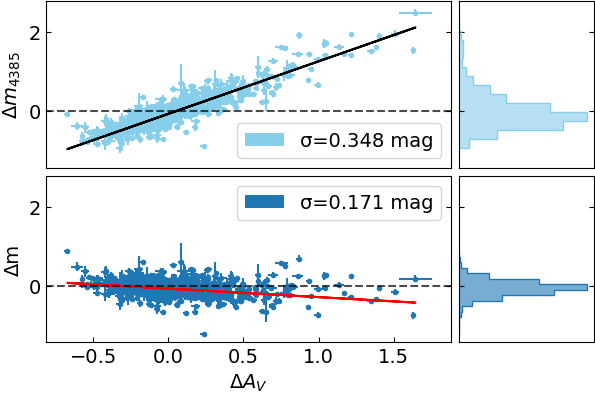}
    \caption{Top: Magnitude offsets for 783 SNe computed at 4385~\AA{} against the RBTL color $\Delta A_V$. Bottom: RBTL offsets, i.e., gray offsets. Right plots: Corresponding distributions. The magnitude dispersions are provided in the legend. The black and red lines highlight the color correlation discrepancy between the two plots.}
    \label{fig:link_RBTL_to_photo}
\end{figure}

We compared $\Delta m$ to a magnitude derived from photometry, as $m_B$. We added a term $\lambda_0$, which we placed at the Bessell-B effective wavelength, at 4385~\AA{}~\citep{bessell1990}, so that the right side of Eq.~\ref{eq:rbtl_model_1} was rewritten for SN $i$ as 
\begin{equation}
    \Delta m_{4385, i} + \Delta {A}_{V,i} \cdot \left[ CL(\lambda) - CL(\lambda_0=4385 \AA)\right],
\end{equation}
with $\Delta m_{4385, i}$ the magnitude offset of the spectrum $f_{\mathrm{max}, i}(\lambda)$, compared to $f_{\mathrm{ref}}(\lambda)$, at 4385~\AA{}\footnote{$\lambda_{0}$ is interpreted here as a leverage wavelength around which the color law rotates with an amplitude depending on the color factor $\Delta A_V$.}. Using the ZTF sample, we computed the RBTL magnitude parameter in the configuration $\Delta m_{4385}$, while the color term $\Delta A_V$ did not change by construction. We show both magnitude distributions in Fig.~\ref{fig:link_RBTL_to_photo}: the dispersion of the 4385~\AA{}-magnitude is 0.347~mag, close to the one expected by photometry, while the $\Delta m$ dispersion is smaller, at 0.171~mag. By plotting both magnitudes against color, we show that the slope is strongly mitigated for RBTL. The next section addresses this remaining correlation with color.

The wavelength range was reduced to 50\% in power (ln) compared to \citetalias{boone2021}, and during the RBTL fit, the weighting was different because we only kept the stable spectral domains: these changes might lead to different RBTL parameters. These impacts would be seen throughout the paper as we compared all ZTF results with those of SNfactory, which we generated under these new conditions.

\subsection{RBTL standardization}
\label{subsec:RBTL_stand}

Similarly to the photometric standardization, we computed a standardization in color of the two SNe samples. To do this, we only considered the $\Delta A_V $ parameter derived from spectroscopy.

\subsubsection{RBTL parameter correlations} 

We followed the quality cut recommendation from \citetalias{boone2021a}, which is summarized Table~\ref{tab:table_cut}.
We only considered SNe with $|\Delta A_V| <0.5$. This is the most significant cut and cuts high-color SNe on which an incorrect $R_V$ would have a high impact. We discarded those at $z<0.03$ to avoid significant effects from peculiar motions. After these cuts, 688 SNe~Ia were left for the ZTF sample and 139 for the SNfactory sample. 
Fig.~\ref{fig:rbtl_salt_color} shows that the cut of $\Delta A_V <0.5$ is similar to the usual $c<0.3$ in cosmology. 

As mentioned Sect.~\ref{subsec:rbtl_applic_method}, $\Delta m$ magnitude corresponds to the gray offset between the (predefined) SN~Ia reference spectrum $f_{\text{ref}}(\lambda)$ relative to spectral observation after the color-law correction. This gray offset is averaged over the spectrum through low-dispersion areas by construction. It can thus be understood as the Hubble residuals, constructed over a bandpass defined by the $\eta(\lambda)<0.073$~mag Boolean mask, taking advantage of spectroscopy to specifically exclude areas of high variability. In addition, we worked at peak brightness as the spectra were rescaled at maximum light. To these Hubble residuals, the associated uncertainties are relative to spectral variations, propagated during the $\Delta m$ fit, mostly due to SNR and spectral uncertainties. 

\begin{figure}
    \centering
    \includegraphics[width=\linewidth]{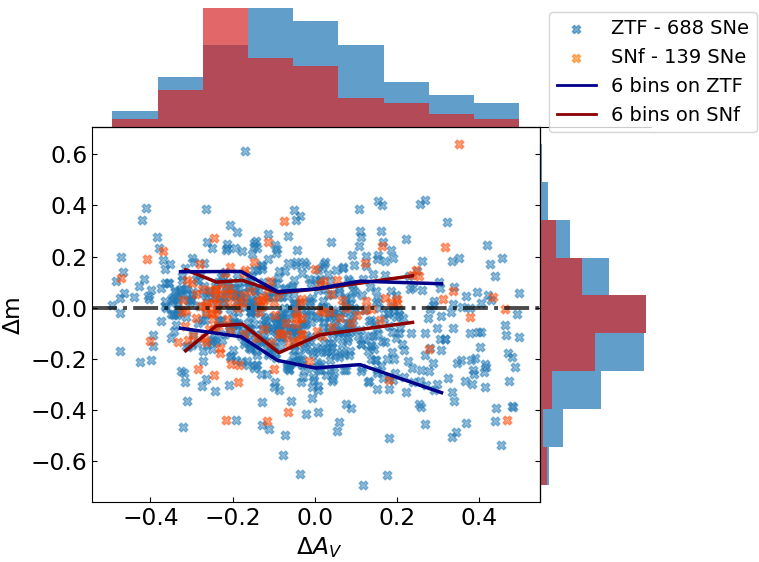}
    \caption{Read-between-the-lines parameter $\Delta m$ as a function of $\Delta A_V$ for ZTF (light blue) and SNfactory (orange). For both samples, $\Delta m$ is binned on $\Delta A_V$ to display the trend with color. It is shown by lines encompassing the uncertainties of the bins for six bins with an equal number of SNe : 114 SNe per ZTF bin (dark blue lines) and 23 per SNfactory bin (red lines). $\Delta m =0$ is indicated as a dashed black line. }
    \label{fig:Dm_versus_Av}
\end{figure}

To verify whether a correlation with $\Delta A_V$ remains after the color-law correction, we compare the residuals $\Delta m$ as a function of $\Delta A_V$ in Fig.~\ref{fig:Dm_versus_Av}. The ZTF sample has a larger number of high $\Delta A_V$ SNe than SNfactory, and these SNe show a broader dispersion in magnitude. By computing the Pearson coefficient for the two samples, we obtained $r=0.076\pm0.085$ for SNfactory and $r=-0.229\pm0.036$  for ZTF, demonstrating that the two samples behave differently in terms of color.

This correlation between the Hubble residuals and $\Delta A_V$ could be absorbed by changing $R_V$, but ZTF and SNfactory present similar population characteristics, and therefore, there is no reason for the color law to be different. We instead fit for two separate slopes to correct for this correlation, and we interpret it in Sect.~\ref{sec:discussion}.

\subsubsection{RBTL standardization method}

Since the $\Delta A_V$-$\Delta m$ correlation is non-negligible, we fit for the linear relation between them with a  $\beta_\mathrm{RBTL}$ slope coefficient. The RBTL Hubble residual of SN $i$ is thus defined as
\begin{equation}
 \Delta \mu_{\mathrm{RBTL}, i} = - \left[\Delta m_i - \beta_\mathrm{RBTL} \cdot\Delta A_{V,i} \right].
 \label{eq:RBTL_Dm}
\end{equation}
We fit $\beta_\mathrm{RBTL}$ by minimizing a total $\chi^2$ , which procedure is described by \cite{ginolin2025a}. It takes the two parameter errors $\sigma_\mathrm{\Delta m, i}$  and $\sigma_\mathrm{\Delta A_{V}, i}$ into account and also the errors on RBTL magnitude and RBTL color of SN $i$.

We obtained $\beta_{\text{RBTL}}= -0.219\pm0.017 $ for ZTF, corresponding to CL($\lambda_0 = 13950 \pm 700$~\AA{}) (see Sect.~\ref{sec:rbtl_residual_scatter}). The $\beta_{\text{RBTL}}$ obtained for SNfactory is $0.096\pm0.039$, which is much closer to zero than ZTF, as expected from Fig.~\ref{fig:Dm_versus_Av}. Simulations matching the c distributions of the two samples showed that the $\beta_{\text{RBTL}}$ discrepancy cannot be explained by the small difference between the photometric color distributions. This rather suggests an incorrect color calibration of the ZTF spectral sample, which affects the redder SNe. This spectral color systematic uncertainty affects the $\Delta m$ errors and causes the increase in scatter. This hypothesis is analyzed in Sect.~\ref{sec:discussion}. When fitting $\beta$ on the volume limited sample (see Sect.~\ref{sec:ztfsniadr2}), we obtained $\beta=-0.203\pm0.022$. We continued with the complete sample to keep good statistics.

Another approach is to fit $\beta_{\text{RBTL}}$ directly on the spectra between the lines instead of on the parameters $\Delta m$ and $\Delta A_V$. The results are consistent, and this approach does not improve the results. Therefore, we continued with the primary approach to fit $\beta_{\text{RBTL}}$ in the $\Delta m$, $\Delta A_V$ plane.

\subsubsection{RBTL Hubble diagram}
\label{subsec:rbtl_hd}

\begin{figure*}
    \centering
    \includegraphics[width=0.7\linewidth]{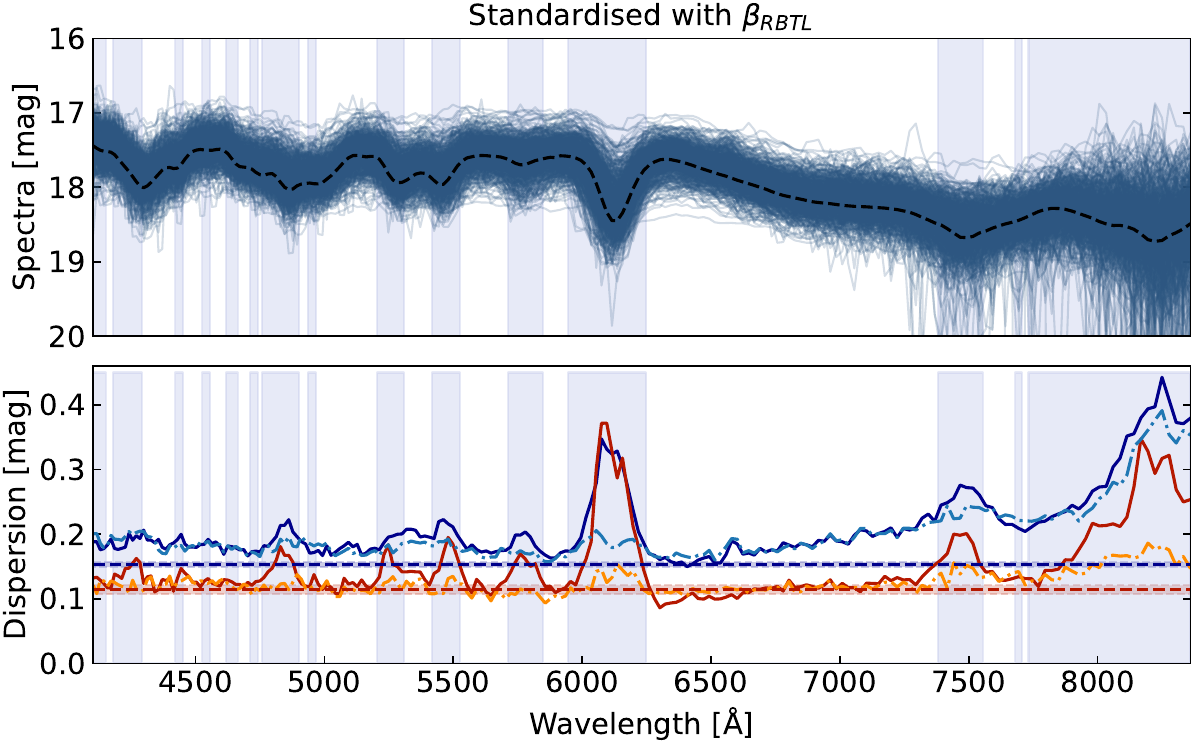}
    \caption{ Top: 688 ZTF spectra in magnitude, after color standardization with $\beta_\mathrm{RBTL}$. The reference flux, derived from SNfactory training, is shown with a dashed black line. Bottom: Dispersions (nMAD) in wavelength for ZTF (blue) and the 139 SNe from SNfactory (red). The dispersion (nMAD) of the scalar residuals is shown with dashed lines at 0.153~mag and 0.114~mag. The residual dispersion in wavelength after an additional line correction for $\vec{\xi}$, described by Eq.~\ref{eq:TE_GP_formula}, is displayed with dashed lines in light blue (ZTF) and light orange (SNfactory).}
    \label{fig:dist_mod_RBTL_stand}
\end{figure*}

After we defined the standardization parameter $\beta_{\text{RBTL}}$, we corrected the spectra at maximum light $f_{\text{max}, i}(\lambda)$ for this standardization in color,
\begin{equation}
    f_{\mathrm{stand}, i}(\lambda)  = f_\mathrm{max, i}(\lambda) \cdot 10^{ +0.4  \cdot  \Delta A_{V, i} \cdot 
    \left[CL(\lambda) - \beta_\mathrm{RBTL}\right]}.
    \label{eq:rbtl_model}
\end{equation}
The standardized spectra of the ZTF sample are shown in Fig.~\ref{fig:dist_mod_RBTL_stand} with the residual dispersion in wavelength for SNfactory and ZTF. We show the dispersion of Hubble residuals, which are defined by Eq.~\ref{eq:RBTL_Dm}, as dashed lines. These dispersions are lower than the spectrum dispersions for both samples because of the SNR variations, which are propagated on the spectral dispersion, but are averaged in the $\Delta m$ dispersion. 
This effect is indeed mitigated for the SNfactory results because their SNR is higher: the ZTF scalar dispersion matches the lowest dispersion values in wavelength, while for SNfactory, it matches the average of the dispersion in wavelength. The absorption line dispersion is again shown to be the main chromatic dispersion error. This highlights the advantage of spectrophotometry to select more standard parts of the spectrum through an averaged gray offset. 

\begin{figure}
    \centering
    \includegraphics[width=\linewidth]{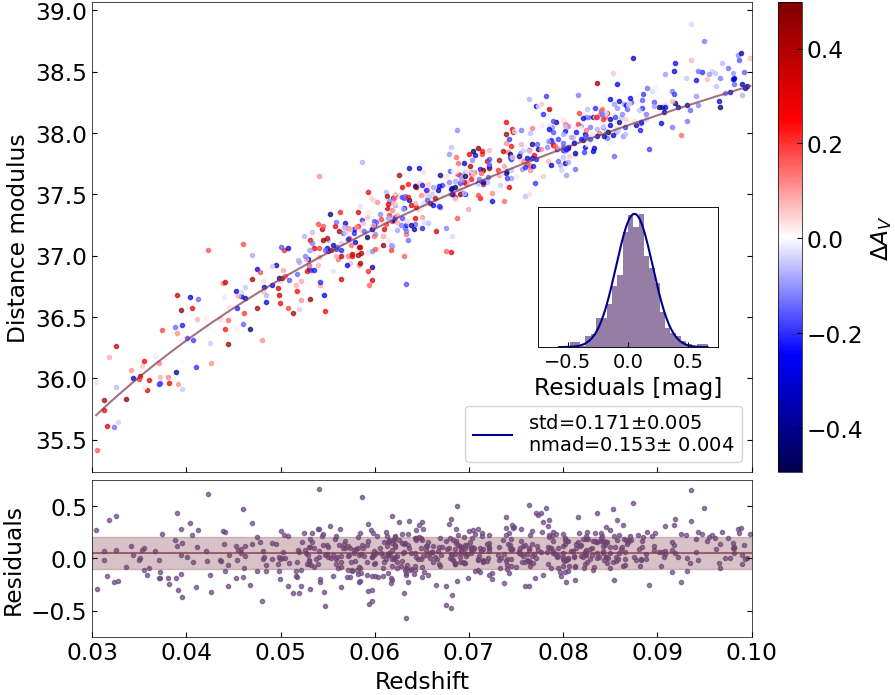}
    \caption{Zwicky Transient Facility Hubble diagram of 688 SNe~Ia (top) and their residuals (bottom), computed between the lines of high variation after a color $\Delta A_V$ standardization. The color bar at the right indicates the $\Delta A_V$ values. The brown lines correspond to the distance modulus values expected by the $\Lambda \text{CDM}$ model from \cite{planck2020} in the top plot, and zeros by definition in the bottom plot.  The residual distribution is shown in the inset, and its dispersion (nMAD) is shown in the legend. This dispersion interval is colored in the bottom plot.}
    \label{fig:standardization_av}
\end{figure}

The resulting Hubble diagram is shown in Fig.~\ref{fig:standardization_av}. The RBTL residual dispersion is $0.153 \pm 0.004$~mag (nMAD) with a single color term, and the STD is $0.171\pm 0.005$~mag. Of this scatter, a $\sim0.071$~mag (nMAD ; $\sim0.078$ STD) contribution is expected from the redshifts fitted on SN features (see Sect.~\ref{sec:discussion}). Another scatter floor of 0.04~mag at least is expected due to spectrophotometric inaccuracy, as explained in Sect.~\ref{subsec:rbtl_applic_method}. We estimated the scatter of the RBTL standardization to be $0.129\pm 0.004$~mag (nMAD) for the ZTF sample of 688 SNe (see the further discussion in Sect.~\ref{sec:discussion}).
By applying the RBTL standardization for the 139 SNe of SNfactory (after standardization cuts), we obtained a scatter of $0.114 \pm 0.007$~mag (nMAD, and $0.145 \pm 0.009$~mag STD). The 0.006~mag difference to the result by \citetalias{boone2021a} comes from the difference in wavelength range discussed in Sect.~\ref{sec:rbtl_residual_scatter}. All results are summarized in Table~\ref{tab:standardization_results}. We conclude that the ZTF data confirm the low Hubble residual scatter associated with this single-parameter spectroscopic standardization, which was first presented using SNfactory data \citep{fakhouri2015, boone2021, boone2021a}.

\begin{table}[]
    \centering
    \caption{ ZTF and SNf Hubble residual dispersion for several analyses in nMAD. }
    \label{tab:standardization_results}
    \begin{tabular}{c|c|c}
                       &    SNf   &  ZTF \\
                       &     (139 SNe)   &   (688 SNe)\\
        \hline                       
        \hline    
        \noalign{\smallskip}                   
       \multicolumn{3}{c}{ RBTL color standardization} \\
       \noalign{\smallskip}
         \hline
        \citetalias{boone2021a}  &  0.108 & \\
        this paper (in 4100-8400 \AA{})& 0.114 & 0.153 \\
        \hline
        Impact of $\sigma_z$ & $\sim$ 0 & 0.071\\
        Impact of flux-calibration & -- & $\ge$0.04 \\
        Host galaxy contamination & -- & $>0$ \\
        \hline
        Corrected dispersion & 0.114 & $\le$0.129\\
        
        \hline
         \hline
        \noalign{\smallskip}
       \multicolumn{3}{c}{ TE standardization} \\
       \noalign{\smallskip}
         \hline
        \citetalias{boone2021a}  &  0.083~\footnote{This dispersion does not take into account the GP error, otherwise the intrinsic dispersion is $0.073$~mag.} & \\
        this paper (in 4100-8400 \AA{})& 0.097& 0.156 \\
        
    \end{tabular}
    \tablefoot{STDs are in the text. }
\end{table}

We obtained a scatter of $0.163\pm 0.004$~mag (nMAD) from these same ZTF SNe~Ia using the SALT standardization, using the standardization parameters values estimated by \citealt{ginolin2025a}. Therefore, the RBTL scatter is lower than the scatter obtained with the photometric standardization, even though it is based on fewer parameters.

\section{Twins Embedding}
\label{sec:te}

After the DTEM and RBTL, the third step to describe SN~Ia dispersion is the manifold reduction introduced by \citetalias{boone2021a} and summarized in Sect.~\ref{sec:twins_embedding_method}. It consists of an Isomap parameterization that captures variabilities in absorption line regions and a GP that uses these variability parameters $\vec{\xi}$ to predict the overall spectral amplitude, that is, to standardize the SN~Ia magnitude. Isomap and GP were both initially trained by \citetalias{boone2021} on 134 selected SNe from SNfactory. 

In this section, we apply the method to the SNfactory dataset (139 SNe~Ia) and the larger ZTF samples (688 SNe~Ia). We first derive the three components of the Isomap in Sect.~\ref{subsec:manifold}, and we test the associated standardization method in Sect.~\ref{subsec:TE_stand}. 

\subsection{Manifold learning: Deriving the Isomap parameters}
\label{subsec:manifold}

The Isomap algorithm we presented in Sect.~\ref{sec:twins_embedding_method} was trained by \citetalias{boone2021} on selected SNfactory data to capture the SN~Ia dispersion that is not considered by the simple RBTL color + magnitude offset model. It typically corresponds to spectral features that were explicitly identified and masked in Sect.~\ref{sec:RBTL} to isolate the color and that contain further information on the SN~Ia variability. 
The Isomap provided by \citetalias{boone2021} was computed on SNfactory data over the 3305--8586~\AA{} wavelength range. To apply it on ZTF data (SEDm), we simply truncated the Isomap vectors to match its 4103--8388~\AA{} wavelength range. 
We also tested retraining the Isomap on SNfactory data while limiting the data to that  wavelength range, and the results were consistent. We therefore used the \citetalias{boone2021} Isomap, which we truncated and applied it to ZTF and SNfactory for self-consistent comparisons.

\begin{figure}
    \centering
    \includegraphics[width=20pc]{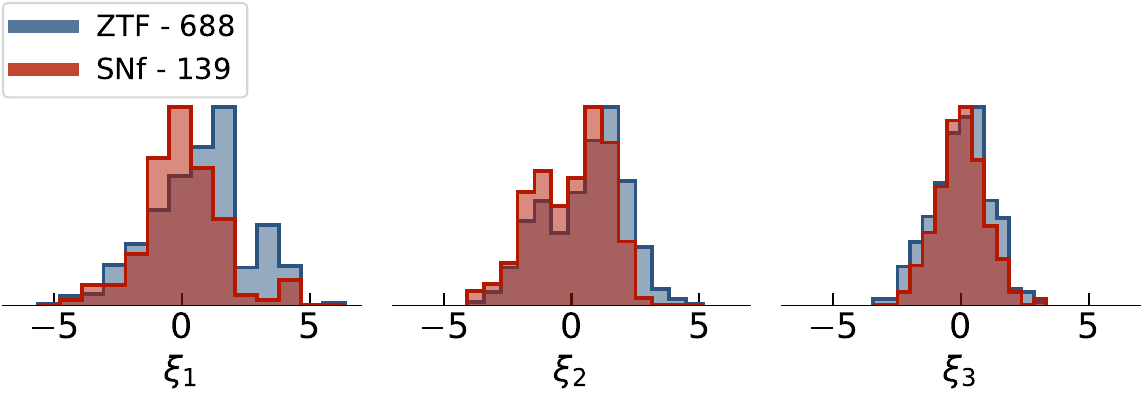}
    \caption{Isomap coordinate distributions for the two standardization samples: ZTF (688 SNe) in blue and SNfactory (139 SNe) in red. }
    \label{fig:xi_with_snf}
\end{figure}

The distributions of the three Isomap parameters $\vec{\xi}_i$ are shown in Fig.~\ref{fig:xi_with_snf}. They agree between the ZTF and SNfactory samples, even though the color-calibration issue potentially affects the ZTF sample, as discussed in Sect.~\ref{sec:RBTL}. This is expected because this Isomap was designed to capture narrow spectral features and is therefore less sensitive to the large-scale calibration issue. We compared the Isomap parameters to SALT stretch $x_1$: as expected, $\xi_2$ is strongly correlated with stretch $x_1$, with a Pearson correlation of $0.59\pm0.025$ for ZTF and $0.83\pm0.026$ for SNfactory.

In addition to the Isomap, \citetalias{boone2021} provided a functionality (based on GPs) to convert Isomap parameters into spectral distortions \footnote{Conceptually, this is similar to converting $x_1$ into its corresponding spectral distortions in the SALT modeling (i.e., $M_1(\lambda)$), except that the \vec{\xi} space is nonlinear.}. 
The resulting spectral modeling is described by both $\Delta A_V$ and $\text{GP}(\lambda)(\vec{\xi})$ in Eq.~\ref{eq:TE_GP_formula}. As illustrated in Fig.~\ref{fig:dist_mod_RBTL_stand}, it captures some of the main SN~Ia chromatic variabilities. However, the two lines that vary most in the initial Isomap training (see Fig.~6 of \citetalias{boone2021}) are not captured with ZTF data: the \ion{Ca}{II}~\text{H\&K} feature is below the wavelength range, and the \ion{Ca}{II}~\text{IR} triplet is not reduced for ZTF, as shown in Fig.~\ref{fig:dist_mod_RBTL_stand}, which might be due to the flux calibration procedure. In addition, the continuum dispersion is not reduced and remains above the RBTL residual dispersion for ZTF and SNfactory: this GP was trained after the full RBTL correction, and model $\eta(\lambda)$. It was therefore not trained to reproduce the overall continuum, which is already captured by $\Delta m$. 

\subsection{Twins-embedding standardization}
\label{subsec:TE_stand}

Given the Isomap parameters $\vec{\xi}$ plus the RBTL color $\Delta A_V$, \citetalias{boone2021a} trained a different Gaussian process to predict the remaining gray offsets (see Eq.~\ref{eq:GP_Dm}). Thus, the residual magnitude $\Delta \mu_{\mathrm{TE}}$, which corresponds to the standardized SN~Ia Hubble residuals of the TE method, can be expressed as 
\begin{equation}
 \Delta \mu_{\text{TE},i} = - \left[\Delta m_i - \beta_\mathrm{RBTL} \cdot\Delta A_{V,i} - \delta m^{\text{GP}}(\vec{\xi}_i) \right].
 \label{eq:GP_Dm}
\end{equation}

\begin{figure}
    \centering
    \includegraphics[width=20pc]{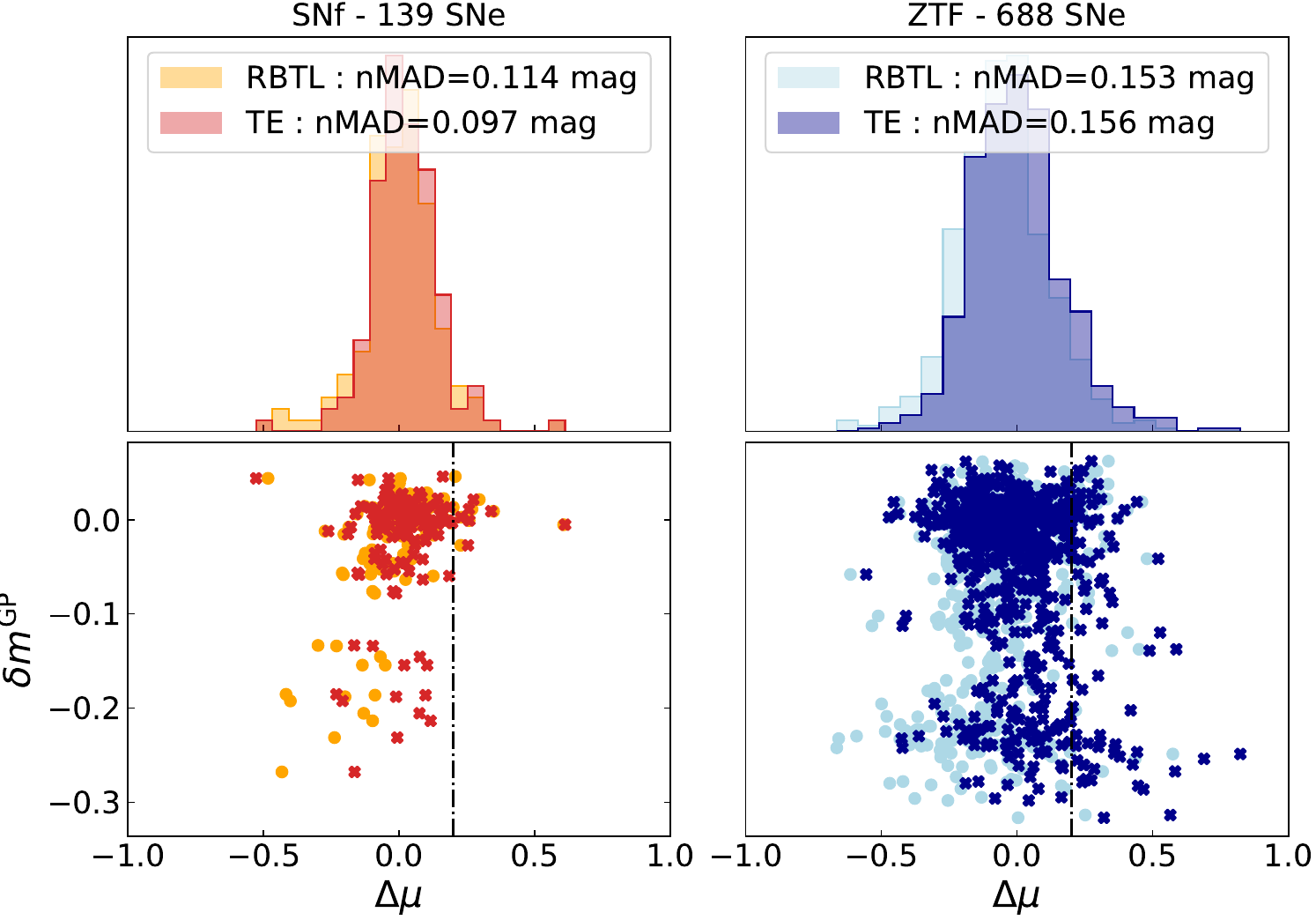}    \caption{Top panels: Histograms of Hubble residuals $\Delta \mu$ after RBTL standardization for SNfactory (left; light orange) and ZTF (right; light blue) and after $\delta m^{\text{GP}}$ correction for SNfactory (dark orange) and  ZTF (dark blue). The dispersions (nMAD) are provided in the legends. Bottom panels: Predicted magnitude $\delta m^{\text{GP}}$ against the magnitude residuals. We highlight the lack of SNfactory data above 0.2~mag with a dashed black line. This region is strongly populated in the ZTF sample.}
    \label{fig:GP_dms}
\end{figure}

We show in Fig.~\ref{fig:GP_dms} the TE Hubble residuals $\Delta \mu_{\text{TE}}$ for SNfactory and ZTF. They are compared with those computed using only the color standardization $\Delta \mu_{\text{RBTL}}$. For SNfactory, the dispersion reduction from $0.114~\mathrm{mag} \pm 0.007$ to $0.097~\mathrm{mag} \pm 0.006$ is smaller than the results from \citetalias{boone2021}, although compatible, as its final dispersion is $0.083 \pm 0.010$~mag. This might be due to the difference in wavelength range, which withdraws the two \ion{Ca}{II} lines. These lines vary most (see \citetalias{boone2021}~Fig.~6.), and they therefore contain most of the variation information. The residual dispersion does not improve for the ZTF sample: the nMAD changes from 0.153~mag to 0.156~mag. This is expected because the GP conversion of $\vec{\xi}$ into magnitude offsets relies on the precision of these Isomap parameters. In addition to the fact that ZTF lacks information on the \ion{Ca}{II} lines, the Isomap requires a high spectral quality, and the spectral uncertainty of the ZTF sample is significant. Figure~\ref{fig:GP_dms} highlights a wider $\Delta \mu_\mathrm{RBTL}$ range for the ZTF, for example, above 0.2, which will not be captured by the training on SNfactory. This correction does not remain for the following steps.

\section{Discussion}
\label{sec:discussion}

In this section, we discuss the consistency of our results.
We study the dependence of the RBTL standardization on SN properties in Sect.~\ref{subsec:dependancies} and its dependence on the host galaxy properties in Sect.~\ref{subsec:Host_corr}. In all results presented in this section, application of the TE standardization with $\vec{\xi}$ only increased the ZTF scatter, even for subsamples identified with a low $\Delta\mu_\mathrm{RBTL}$ scatter. We hence focused on studying the origin of the increased variance found in ZTF in comparison to SNfactory.

\subsection{Parameter dependences}
\label{subsec:dependancies}

\begin{figure}
    \centering
    \includegraphics[width=\linewidth]{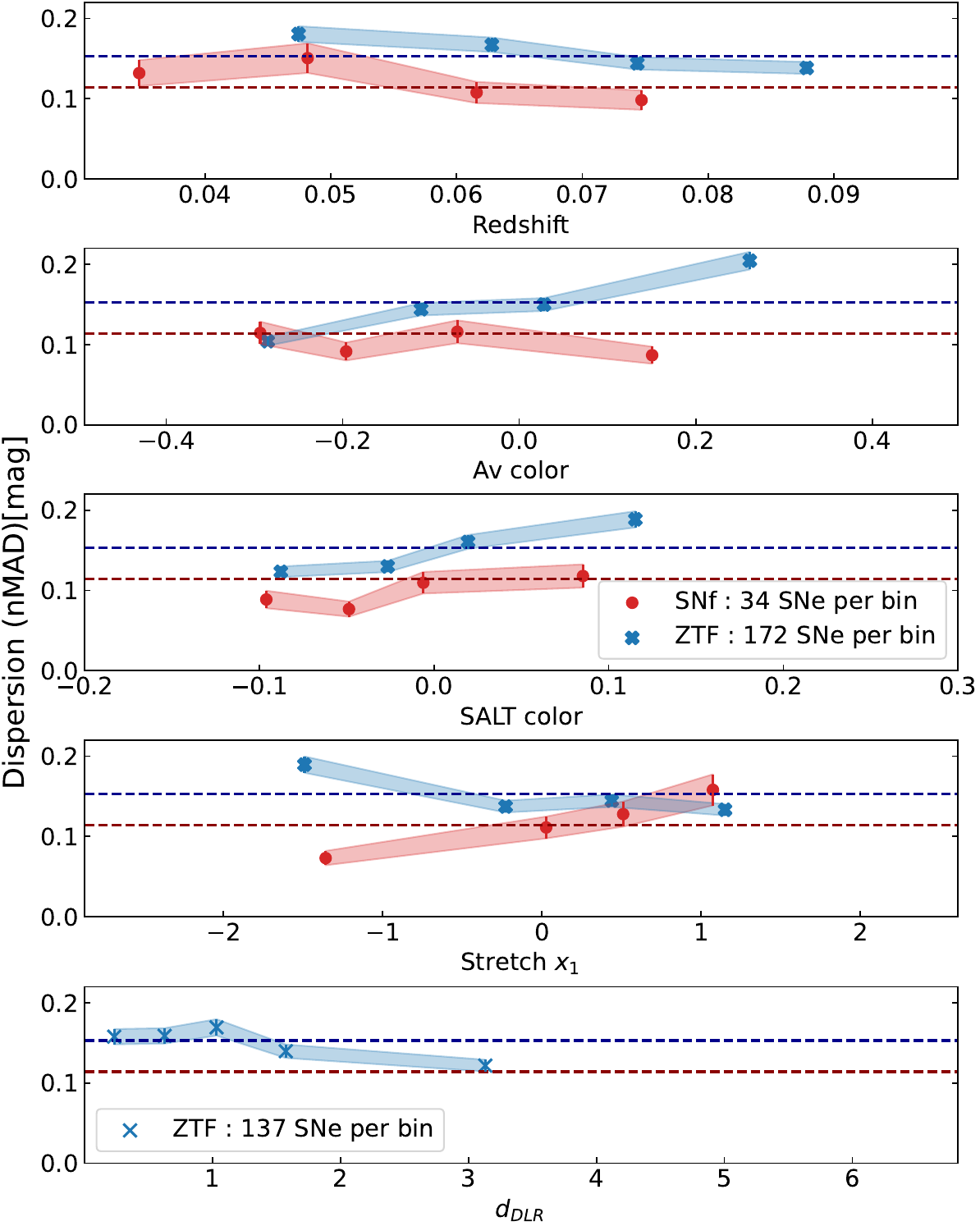}   
    \caption{Read-between-the-lines residual dispersion (nMAD) binned into four bins on the number of SNe as a function of (top to bottom) redshift, RBTL color $\Delta A_V$, SALT color $c$, SALT stretch $x_1$, and $d_{DLR}$ for the ZTF in blue, with 172 SNe per bin (or 5 bins of 137 SNe for $d_\mathrm{DLR}$ parameter), and SNfactory in red, with 34 SNe per bin. The $x$-axis limits are those of the underlying parameters. The RBTL standardization dispersion for the full samples is shown with dashed lines for ZTF (0.153~mag in blue) and for SNfactory (0.114~mag in red).}
    \label{fig:res_dependency}
\end{figure}

To investigate the RBTL standardization dependences on  parameters such as the SN~Ia redshift, the RBTL color, the LC stretch and color parameters, and the distance to the host galaxy, we binned our data into four bins with an equal sample size: 173 SNe~Ia per bin for ZTF, and 34 for SNfactory. The $\Delta\mu_\mathrm{RBTL}$ dispersion (nMAD) as a function of the parameters is shown in Fig.~\ref{fig:res_dependency}.

First, ZTF SNe~Ia with a negative stretch have a larger scatter ($0.189\pm 0.01$~mag) than those with $x_1>0$ ($0.138\pm0.007$~mag). This difference likely arises because the stretch parameter was not used in the DTEM (see Sect.~\ref{sub:te_phase_correction}): the DTEM has a higher dispersion for low-stretch SNe~Ia especially, as shown in Appendix~\ref{app:DTEM_with_stretch}. The SNfactory dataset is less affected because most SNe~Ia have at least one spectrum close to maximum light. Additional work on the DTEM might reduce this stretch dependence. 

Second, the residual scatter increases along with the spectroscopic color $\Delta A_V$, and the bluest ($\Delta A_V<-0.2$) bin has a scatter of $\sim0.1$~mag, which is as low as that from SNfactory. We showed in Sect.~\ref{subsec:RBTL_results} that the $\Delta A_V$ distribution is significantly redder for ZTF than for SNfactory, unlike that of SALT $c$. This suggests that a flux calibration problem causes the spectra to redden, as a left-over host galaxy signal contamination in SEDm spectra despite the \texttt{Hypergal} procedure \citep{lezmy2022a}. This is supported by Appendix~\ref{app:high_slope_calibration}, which shows that the SNe whose color differs depending on SALT or RBTL, the latter being redder, correspond to those with the largest calibration slope correction ($a_1>0.5$). These SNe are shown in Fig.~\ref{fig:dlr_coefs} as corresponding to be close to the host galaxy, with $d_\mathrm{DLR}<1$. This systematic color error would result in a poor fit for $\Delta m$, which would consequently add dispersion to the residuals.

Third, the scatter decreases as the $d_\mathrm{DLR}$ increases, which supports the hypothesis that leftover host contamination scatters the residuals. No significant dependence on redshift is detected. As discussed in Sect.~\ref{subsec:rbtl_hd} and illustrated in Fig.~\ref{fig:correction_rbtl}, we expect no dependence on SNR as it is absorbed by $\Delta m$. 

Finally, we studied the dependence on the redshift measurement method because $\sim43\%$ of the SN redshifts were derived from SN features, with a precision of $\sigma_z=3\times10^{-3}$ (see \citealt{rigault2025}). We show in Appendix~\ref{app:z_source_res} that for a given sample, using these redshifts instead of galaxy redshifts adds a residual scatter of $\sim0.1$~mag. In our sample, the residual scatter is nevertheless lower for the SNe with low-quality redshifts because they are bluer on average. Therefore, instead of removing these SNe, we estimated the additional scatter due to those redshifts by taking each SN redshift error into account. We found a scatter (nMAD) of $~0.071$~mag.

\subsection{Host correlations}
\label{subsec:Host_corr}

We finally studied the correlations between the Hubble residuals $\Delta\mu_\mathrm{RBTL}$ and host galaxy properties. Using ZTF SN~Ia DR2 data, \cite{ginolin2025a} reported a large magnitude offset, also known as the step, between the SALT2 standardized SN~Ia magnitude split as a function of their global host stellar masses or their local environmental colors $(g-z)$, defined within 2~kpc around the SN. When they are measured after stretch and color standardization, these steps are $\sim0.1$~mag, and they are higher when they are fitted simultaneously (see details in \citealt{ginolin2025a}).

\begin{figure}
     \centering
     \includegraphics[width=\linewidth]{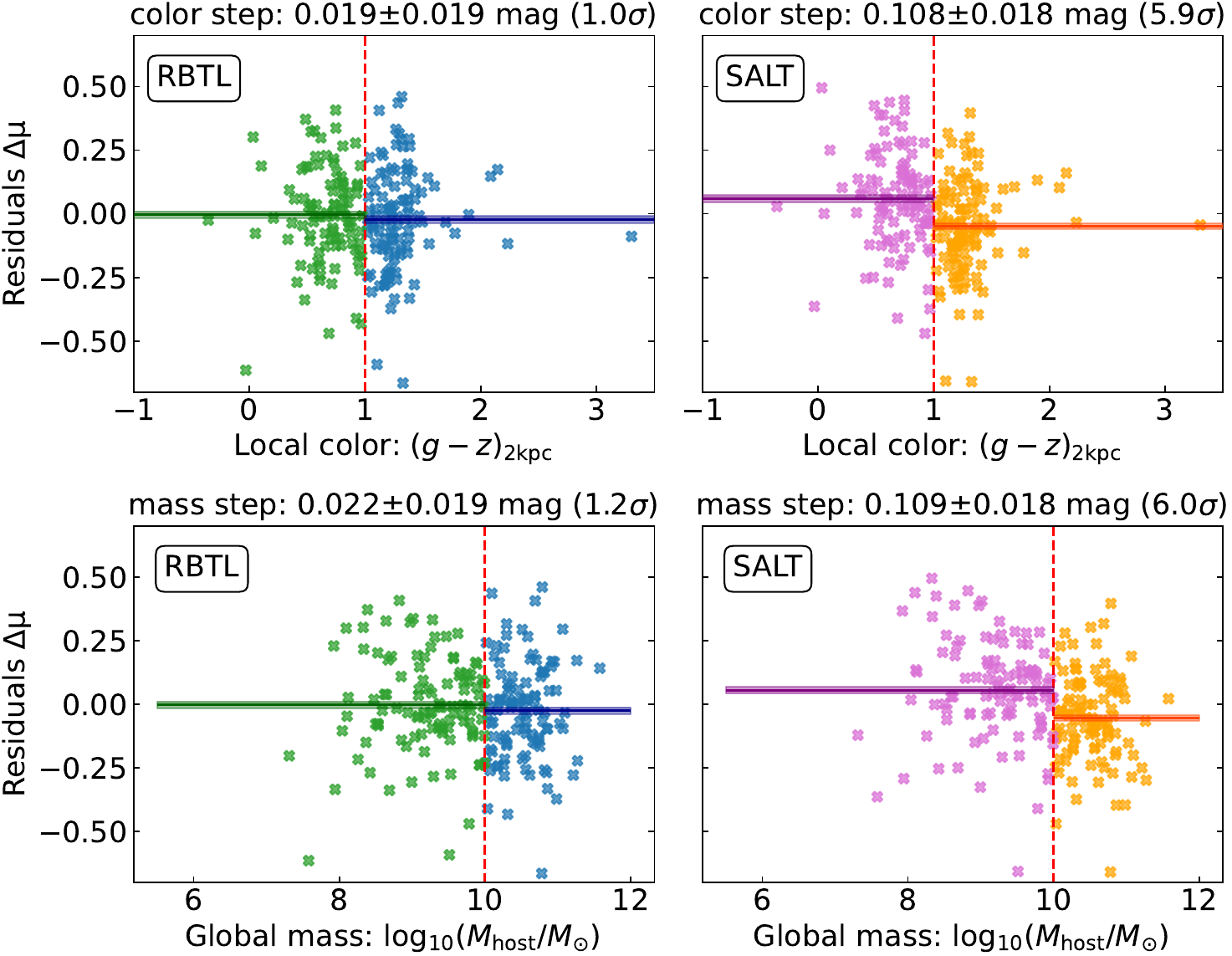}
     \caption{Local color step (top panels) and global mass step (bottom panels) for the volume-limited 209 SNe (of 663) of the ZTF sample after RBTL standardization (left panels) and for the SALT residuals (right panels). The dashed red lines separate the two populations at $(g-z)=1$ and $M_{\text{host}}=10^{10}M_{\odot}$. The means of the different populations and their uncertainties are shown by plain lines.}
     \label{fig:steps}
 \end{figure}

We replicated these results using the ZTF spectrophotometric subsample and compared the step amplitude after the usual stretch and color standardization (as provided by \citealt{ginolin2025a}) and after the RBTL standardization alone. As in \citealt{ginolin2025a}, we limited our study to the ZTF volume-limited sample ($z<0.06$, 219 out of the 688 ZTF SNe~Ia) to cancel selection function effects, and we split the sample at $M_{\text{host}}=10^{10}M_\odot$ and $(g-z)_\mathrm{2kpc}=1$ to define low and high host stellar mass or local color environments, respectively.
The results are shown in Fig.~\ref{fig:steps}.

We recovered the steps of \cite{ginolin2025a} using our subsample and found the SALT2 photometric stretch and color standardization with a leftover $~0.11$~mag environmental step (similar for both environment tracers). However, the amplitude of the astrophysical bias was strongly reduced using the $\Delta A_V$ RBTL spectroscopic standardization, to $0.019\pm 0.019$~mag for the local color and to $0.022\pm0.019$~mag for the global host mass. The two steps are two to three times smaller than their photometrically standardized counter part, and they are altogether only measured at the $\sim1.2\sigma$ level. These results are consistent with those previously reported by \citetalias{boone2021a} using the SNfactory dataset. The host steps would be caused by the variability of strong spectral features, which are ignored by the RBTL fit.

When the volume-limited cut was removed, the steps were more pronounced: the local color step was $0.036\pm 0.010$ for RBTL residuals, and it was $0.098\pm 0.009$ for the SALT standardization; and the mass step was $0.036\pm 0.010$~mag for the RBTL residuals, and it was $0.101\pm 0.009$ for the SALT standardization. 

\section{Conclusion}
\label{sec:conclusion}

The SEDm was designed as a typing spectrograph rather than a spectrophotometric instrument. Therefore, we first calibrated the spectra of the SN Ia DR2 sample in flux using the associated $g,r,i$ photometric LCs, calibrated at the percent level. We computed the synthetic photometry of the spectra, compared them to the photometric data interpolated in time by SALT, and then corrected the spectra for a polynomial of second order. We estimated the final flux dispersion floor at the level of $\sim~0.07$~mag, with a contribution of a flux-calibration residual of $\sim~0.04$~mag and a contribution from low SNR. We release the final sample of 1897 flux-calibrated spectra from 1607 SNe Ia.

We have presented the first application of the TE spectrophotometric standardization method to a dataset other than the SNfactory sample. The ZTF subset contained 783 SNe after we applied the quality cuts, corresponding to about four times the SNfactory sample. The first step of TE is a spectral correction for the phase, and we estimated its accuracy using the ZTF sample to be at 0.011~mag in $g$ band. The second step, fitting a color $\Delta A_V$ per SN on stable spectral domains, presented redder SNe for ZTF sample. A first standardization, based on this RBTL color parameter $\Delta A_V$ alone, resulted in a dispersion of 0.153~mag for the ZTF sample and of 0.114~mag for the SNfactory sample. The resulting dispersion for the ZTF sample falls at 0.129~mag when the contributions of redshift error and spectrophotometric error are taken into account. In addition, ZTF $\Delta A_V$ are significantly redder than SNfactory, while their photometric color distributions are similar. The $25\%$ bluest ZTF SNe and the $20\%$ farthest from the host galaxies reached an SNfactory dispersion of $\sim0.1$~mag, indicating that the remaining host galaxy contamination within the ZTF data contributes to increase the scatter. 

While the RBTL standardization was based on only one parameter, the SALT photometric standardization reached a dispersion of 0.164~mag for the same SN sample for the stretch and color correction and a dispersion of $\sim0.151$~mag with the additional step correction. We computed host steps for the RBTL residuals, and the mass step and local color steps were both consistent with zero. Thus, a cosmological analysis using the RBTL method would be less affected by astrophysical biases than an analysis using SALT.

The third step of the TE is a nonlinear parameterization of chromatic variations based on three parameters $\vec{\xi}_{1,2,3}$. The results for the ZTF spectral dataset were sufficient to explain some line variations, but not the two main lines that vary most in the TE training. The standardization based on these nonlinear TE parameters, which predicts the magnitude offset $\Delta m$ per SN, is not effective on this dataset: this was expected as the instrumental error (SNR, galaxy subtraction, and flux calibration) of the ZTF spectrum dataset is large.

The Nancy Grace Roman Space Telescope, scheduled for launch in 2026, has a resolution prism of $R~\sim~100$ and a wavelength range of 7500--18\,000~\AA{}. The wider wavelength range is interesting as the dust will not contaminate  the signal in the infrared, and it reaches the low dispersion domain around 7000~\AA{} for redshifts up to $z=1.6$. The presence of atmospheric water bands, perceived by the ground-based telescopes of ZTF and SNfactory, while mitigated in the case of SNfactory, are the main contaminant of the signal at wavelengths above 7500~\AA{}. When we focus on galaxy subtraction and redshift quality, the RBTL standardization of the spectral sample would result in a Hubble residual scatter of $0.129$~mag (upper limit). In addition, the full TE standardization could be applied if the spectral data present a good spectral quality, which would further reduce the Hubble residual dispersion.

\begin{acknowledgements}
Based on observations obtained with the Samuel Oschin Telescope 48-inch and the 60-inch Telescope at the Palomar Observatory as part of the \textit{Zwicky} Transient Facility project. ZTF is supported by the National Science Foundation under Grant No. AST-1440341 and a collaboration including Caltech, IPAC, the Weizmann Institute of Science, the Oskar Klein Center at Stockholm University, the University of Maryland, the University of Washington, Deutsches Elektronen-Synchrotron and Humboldt University, Los Alamos National Laboratories, the TANGO Consortium of Taiwan, the University of Wisconsin at Milwaukee, and Lawrence Berkeley National Laboratories. Operations are conducted by COO, IPAC, and UW.
SED Machine is based upon work supported by the National Science Foundation under Grant No. 1106171.
The ZTF forced-photometry service was funded under the Heising-Simons Foundation grant \#12540303 (PI: Graham).
This project has received funding from the European Research Council (ERC) under the European Union’s Horizon 2020 research and innovation programme (grant agreement n 759194 - USNAC).
KM, U.B and T.E.M.B. acknowledges funding from Horizon Europe ERC grant no. 101125877.
This work has been supported by the research project grant “Understanding the Dynamic Universe” funded by the Knut and Alice Wallenberg Foundation under Dnr KAW 2018.0067 and the {\em Vetenskapsr\aa det}, the Swedish Research Council, project 2020-03444.
L.G. acknowledges financial support from AGAUR, CSIC, MCIN and AEI 10.13039/501100011033 under projects PID2023-151307NB-I00, PIE 20215AT016, CEX2020-001058-M, ILINK23001, COOPB2304, and 2021-SGR-01270.
Y.-L.K. was supported by the Lee Wonchul Fellowship, funded through the BK21 Fostering Outstanding Universities for Research (FOUR) Program (grant No. 4120200513819) and the National Research Foundation of Korea to the Center for Galaxy Evolution Research (RS-2022-NR070872, RS-2022-NR070525).
G.A. and S.P. were supported in part by the Director, Office of Science, Office of High Energy Physics of the U.S. Department of Energy under Contract No. DE-AC025CH11231.
MG is supported by the European Union’s Horizon 2020 research and innovation programme under ERC Grant Agreement No. 101002652 (BayeSN; PI K. Mandel).

  In the development of our pipeline, we acknowledge use of the Python 
  libraries Numpy 
  \citep{harris2020}, Scipy 
  \citep{virtanen2020},
  Matplotlib 
  \citep{hunter2007}, Astropy 
  \citep{astropycollaboration2013, astropycollaboration2018, astropycollaboration2022} and
  Pandas 
  \citep{thepandasdevelopmentteam2025}.

\end{acknowledgements}

\bibliographystyle{aa}
\bibliography{Bibliography.bib}

\begin{appendix}

\section{RBTL spectral dispersion}
\label{app:RBTL_spectral_disp}

\begin{figure}
    \centering
    \includegraphics[width=\linewidth]{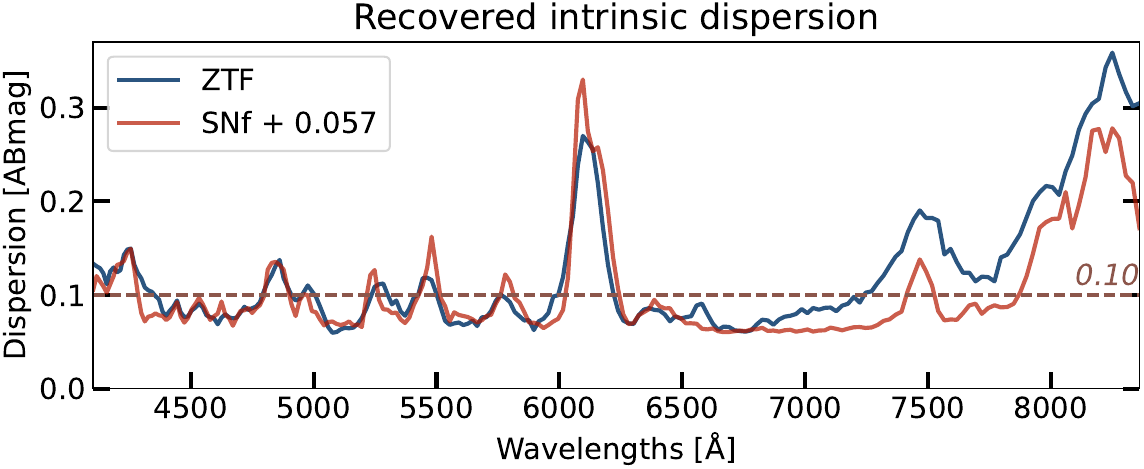}
    \caption{Spectral dispersion after correction for both RBTL parameters, for both 783 SNe of ZTF (dark blue) and 200 SNe of SNfactory (dark red), to which we added a 0.057~mag dispersion in quadrature.}
    \label{fig:correction_rbtl_007}
\end{figure}

To compare the residual dispersion in wavelength of SNfactory and ZTF after RBTL correction, we add the ZTF dispersion floor due to the larger SNR, quantified in Sect.~\ref{sec:calib}, of 0.057~mag to SNfactory spectral dispersion, in quadrature, as shown in Fig.~\ref{fig:correction_rbtl_007}. 
We can see that SNfactory dispersion is comparable to the ZTF one, below about 7000~\AA{}. 

\section{Redshift source impact on standardization scatter}
\label{app:z_source_res}

\begin{figure}
    \centering
    \includegraphics[width=0.8\linewidth]{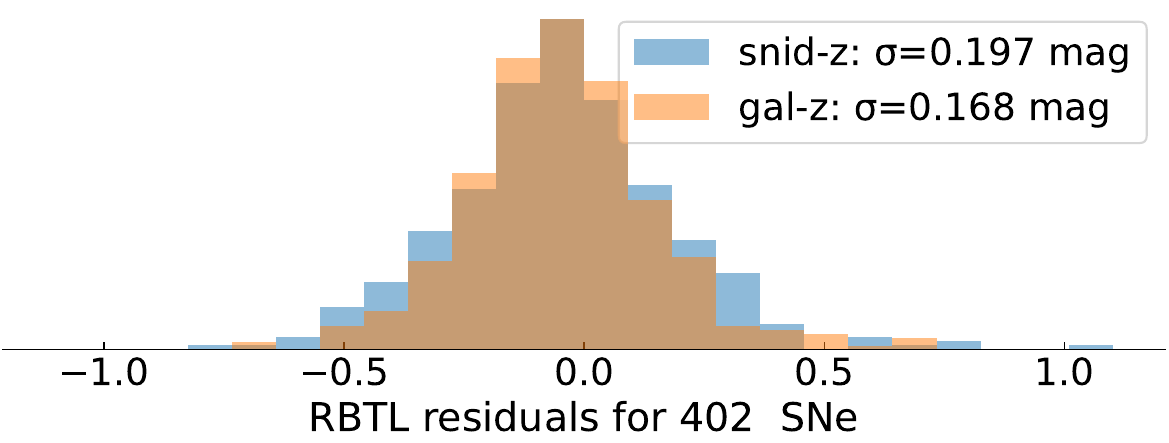}
    \caption{RBTL residual distributions for the same 402 SNe, using either their gal-z (orange) or their snid-z (blue).}
    \label{fig:z_source_res}
\end{figure}

Redshift uncertainty depends strongly on the redshift source: there is either a high-quality ($\sim10^{-5}$) redshift derived from host galaxy, which we label gal-z, or a lower quality redshift ($\sim10^{-3}$) derived from SN features, labeled snid-z. The sample of 783 SNe on which we computed the RBTL standardization includes 402 SNe for which we have access to both redshift sources. To estimate the impact of the redshift quality on the Hubble residuals, we compute the RBTL standardization of this subsample, using either the gal-z redshifts or the snid-z redshifts. We show both residual distributions in Fig.~\ref{fig:z_source_res}: their dispersions (nMAD) are 0.168~mag (gal-z) and 0.197~mag (snid-z). This dispersion discrepancy corresponds to a difference of 0.1~mag, in quadrature.

\section{DTEM with stretch correction}
\label{app:DTEM_with_stretch}

 In practice, \citetalias{boone2021} assumes $c'_1(\lambda)=c'_2(\lambda)=0$ to make  DTEM independant of photometric data. We now apply the DTEM taking into account the stretch to the 783 ZTF SNe:
\begin{equation}
\begin{aligned}
    m_i(\lambda,p) - m_i(\lambda, 0) &= \left(c_1(\lambda) \cdot p + c_2(\lambda) \cdot p^2\right) \\
   &+ x_1 \cdot \left(c'_1(\lambda) \cdot p + c'_2(\lambda) \cdot p^2\right),
\end{aligned}
\label{eq:max_light_formula_x1}
\end{equation}
with the vectors $c'_1(\lambda)$ and $c'_2(\lambda)$, that corrects for the stretch $x_1$, provided by the SN~Ia LCs.

\begin{figure}
    \centering
    \includegraphics[width=0.8\linewidth]{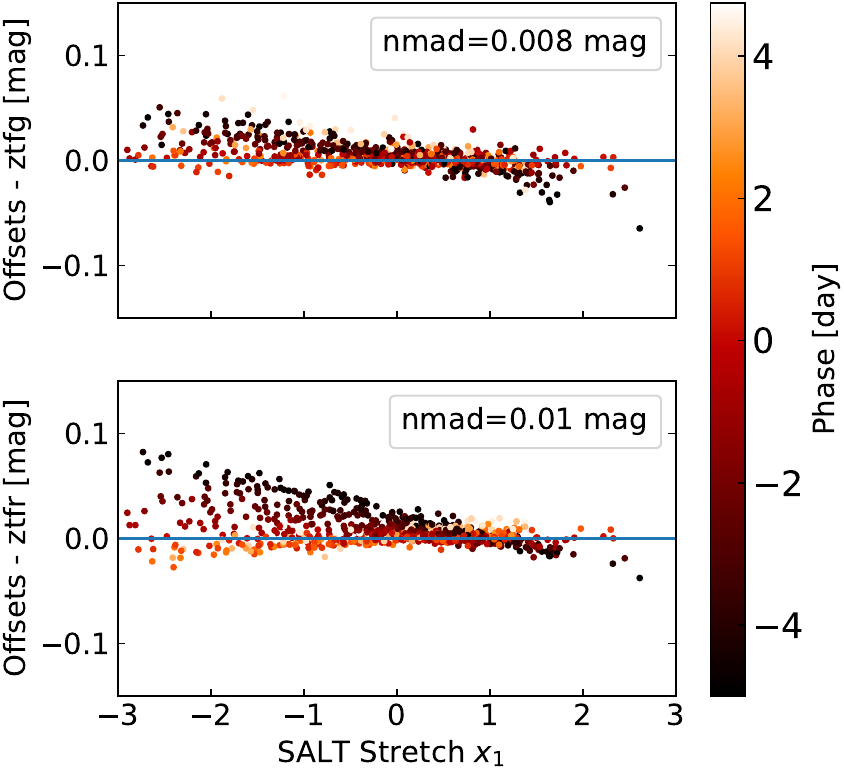}
        \caption{Magnitude offsets for 783 ZTF SNe, between LCs at peak and integrated ZTF spectra corrected for DTEM with stretch correction, in both $g$ (top plot) and $r$ (bottom plot), as a function of the SN stretch $x_1$. The colorbar indicates the initial phase of the spectrum. A blue line indicates the 0 value. The offsets nMAD are given in the legends.}
    \label{fig:DTEM_x1}
\end{figure}

 We integrated the corrected spectra of 783 SNe through the ZTF camera filters, as in Sect.~\ref{sub:te_phase_correction}, and compared the obtained photometric fluxes with those of the LCs at peak. Figure~\ref{fig:DTEM_x1} shows the offsets with the LC, in the $g$ and $r$ filters, as a function of the stretch $x_1$, and the colorbar indicates the phase. The model does not properly correct the spectra for the phase, as the offsets are not exactly zero. We see a strong correlation with the stretch, especially in $r$, and for low stretch. As shown in Fig.~\ref{fig:rbtl_salt_color}, SNfactory has a few low-stretch (i.e., fast) explosions; therefore, the DTEM could be less effective for those SNe. It can also be due to the quadratic evolution not being sufficient to catch all the information of fast SNe. This effect could also be reduced with the $\xi_2$ parameter, shown to be strongly correlated with $x_1$ (see Sect.~\ref{sec:te}).

\section{Calibration residual effects on color}
\label{app:high_slope_calibration}

\begin{figure}
    \centering
    \includegraphics[width=\linewidth]{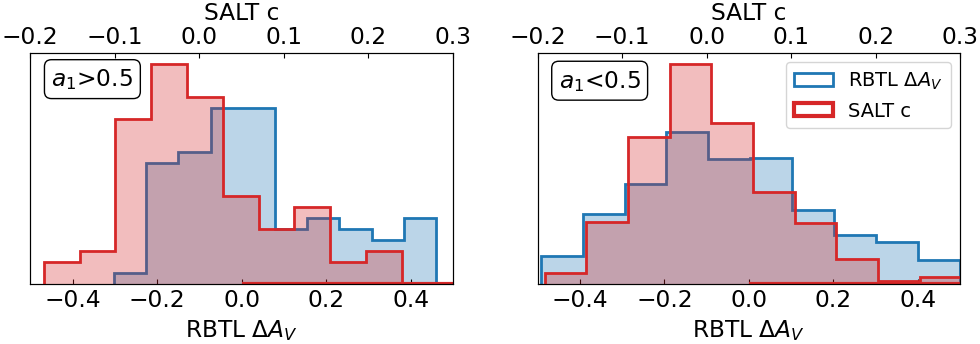}
    \caption{SALT color (red) and RBTL color (blue) distributions of 82 SNe with a slope coefficient $a_1>0.5$ (left) and 600 SNe with $a_1<0.5$ (right).}
    \label{fig:high_calibration}
\end{figure}

We show in Fig.~\ref{fig:high_calibration} the distributions of both RBTL color and SALT color for the 82 ZTF SNe with large flux-calibration correction in slope, that we define with $a_1>0.5$, and 600 ZTF SNe with $a_1<0.5$. Spectra with large slope corrections display redder RBTL color than SALT color. This suggests that the excess of high-$\Delta A_V$ SNe in ZTF could be due to the flux calibration, which adds a systematic error in color.

\end{appendix}

\end{document}